\documentclass[journal]{IEEEtran}
\usepackage{graphicx}
\graphicspath{{../pdf/}{../jpeg/}}
\DeclareGraphicsExtensions{.pdf,.jpeg,.png}
\usepackage{cite}
\usepackage[hidelinks]{hyperref}
\usepackage{enumerate}
\usepackage{subfig}
\usepackage{stfloats}
\usepackage{tabularx}
\usepackage{booktabs}
\usepackage{amsmath,bm}
\usepackage{amsfonts}
\usepackage{amssymb}
\usepackage{color}
\usepackage{multirow}
\begin{document}
	
\title{Tire Slip Angle Estimation based on the Intelligent Tire Technology}	
	
\author{
		Nan Xu, Yanjun Huang, Hassan Askari, and Zepeng Tang  
	
\thanks{This work has been submitted to the IEEE for possible publication. Copyright may be transferred without notice, after which this version may no longer be accessible. This research is supported by National Natural Science Foundation of China (Grant Nos.51875236 and 61790561), China Automobile Industry Innovation and Development Joint Fund (Grant Nos. U1664257 and U1864206)}
\thanks{N. Xu is with the State Key Laboratory of Automotive Simulation and Control, Jilin University, Changchun, Jilin, 130025, China and also with the Department of Mechanical and Mechatronics Engineering, University of Waterloo, ON. N2L3G1, Canada, e-mail: (xunan@jlu.edu.cn).}
\thanks{Y. Huang and H. Askari are the Department of Mechanical and Mechatronics Engineering, University of Waterloo, ON. N2L3G1, Canada e-mail: (y269huan@uwaterloo.ca and askari.waterloo@gmail.com)}
\thanks{Zepeng Tang is with the State Key Laboratory of Automotive Simulation and Control, Jilin University, Changchun, Jilin, 130025, China)}}
	
\maketitle	
	
\begin{abstract}
Tire slip angle is a vital parameter in tire/vehicle dynamics and control.~This paper proposes an accurate estimation method by the fusion of intelligent tire technology and machine-learning techniques. The intelligent tire is equipped by MEMS accelerometers attached to its inner liner.~First, we describe the intelligent tire system along with the implemented testing apparatus. Second, experimental results under different loading and velocity conditions are provided.~Then,~we show the procedure of data processing, which will be used for training three different machine learning techniques to estimate tire slip angles.~The results show that the machine learning techniques, especially in frequency domain, can accurately estimate tire slip angles up to 10 degrees.~More importantly, with the accurate tire slip angle estimation, all other states and parameters can be easily and precisely obtained, which is significant to vehicle advanced control, and thus this study has a high potential to obviously improve the vehicle safety especially in extreme maneuvers. 
\end{abstract}	
	
	
{}

\section{INTRODUCTION}
	
Vehicle dynamics and systems have been developed for several decades in terms of component, modeling, control, and more recent topics e.g. the Intelligent Transportation System (ITS) aspects \cite{mastinu2014road}, autonomous driving \cite{huang2019novel}, and connected vehicles\cite{he2017integrated}.~As the core of the vehicle dynamics and to achieve different vehicle-control purposes, tire dynamics also has been extensively studied.~Indeed, several thorough analysis have been done by researchers in the area of tire dynamics, which resulted in  the Fiala model (1950s) \cite{efiala1954Seitenkrafte}, Brush model \cite{pacejka1991shear}, Dugoff \cite{dugoff1970analysis} (1970s), Magic Formula \cite{pacejka1992magic}, UniTire \cite{guo2005unitire}, LuGre \cite{de1995new}, and Ftire model \cite{gipser2007ftire} (1990s).~Generally, tire model, no matter the theoretical, semi-empirical, or pure empirical, is a set of equations that maps the tire operating parameters e.g. slip rate to the general forces acting on the tire. 

Almost all the vehicle control actions work through the tire forces, which should be accurately acquired to achieve the satisfactory control performance \cite{ahmadi2009adaptive}, \cite{zhao2018control}.~Due to the technical or economic reasons, tire forces with other parameters can only be estimated \cite{jin2019advanced}.~More precisely, in current vehicle control (e.g. lateral stability control), tire forces, vehicle sideslip angle, and wheel cornering stiffness are estimated by using measurements (yaw rate, longitudinal/lateral accelerations, steering angle and angular wheel velocities) only from sensors available to modern cars \cite{baffet2009estimation}. The existing literature uses different estimation methods such as sliding mode observer, Kalman Filter \cite{cheli2007methodology}, extended Kalman filter \cite{baffet2008experimental}, unscented Kalman filter \cite{zhang2017hierarchical} in a centralized or hierarchical manner. 

These observers adopt either kinematic or dynamics vehicle model with simple tire models e.g. linear, linearized, or nonlinear version of aforementioned models \cite{acosta2018virtual}. However, under combined or extreme driving conditions, such simple models are not able to represent the real behaviors of the vehicle due to the fact that tire and its interaction forces with road change quickly and perhaps significantly.~Therefore, the observers fails to perform well because the parameter changes usually faster than the convergence rate of the estimation algorithms \cite{erdogan2011tire}.~In addition, most of the stability control systems integrate noisy and biased measurements from on-board sensors but such noises or inaccuracy of each sensor used in the estimation algorithms are increased under extreme conditions \cite{bevly2004determination}.

It is known that obtaining reliable tire forces  is extremely important to vehicle control performance, and accordingly, to the safety of the drivers and passengers\cite{zhao2018control,naets2017design}.~Therefore, the direct or indirect estimation methods based on intelligent tire technologies can lead to promising results because such technologies can obtain more accurate tire variables by using the measurements e.g. deformation or strain from sensors attached to the tire. Furthermore, it is expected that the control algorithm associated with intelligent tire technology be more effective and simpler since the uncertainties in the observer
will be eliminated and the estimation methods will be largely simplified \cite{lee2017intelligent} \cite{braghin2006measurement}.

Different types of sensors have been used in the intelligent tire studies, such as surface acoustic wave sensor, capacitive sensor, hall-effect sensor, magnetic tire sidewall deflection sensor, nanogenerators\cite{askari2018flexible}, optical sensor, accelerometers \cite{matsuzaki2015intelligent}.~Authors in \cite{lee2017intelligent} present different types of the intelligent tires and the associated measured variables.~A comprehensive review on tire forces estimation and sensing techniques can be found in \cite{askari2019tire} and \cite{audisio2010birth}.~Reference \cite{erdogan2009novel} discusses the pros and cons of each intelligent tire technology.~Among those sensors,  accelerometer is the most popular one for intelligent tire applications,  as mentioned in \cite{erdogan2011tire}.~As a result, this study also implements MEMS accelerometers to extract the acceleration data in three directions, which is then utilized to estimate tire slip angles.

\begin{figure}[!htb]
	\centering
	\vspace{-1em}
	\includegraphics[width=\linewidth]{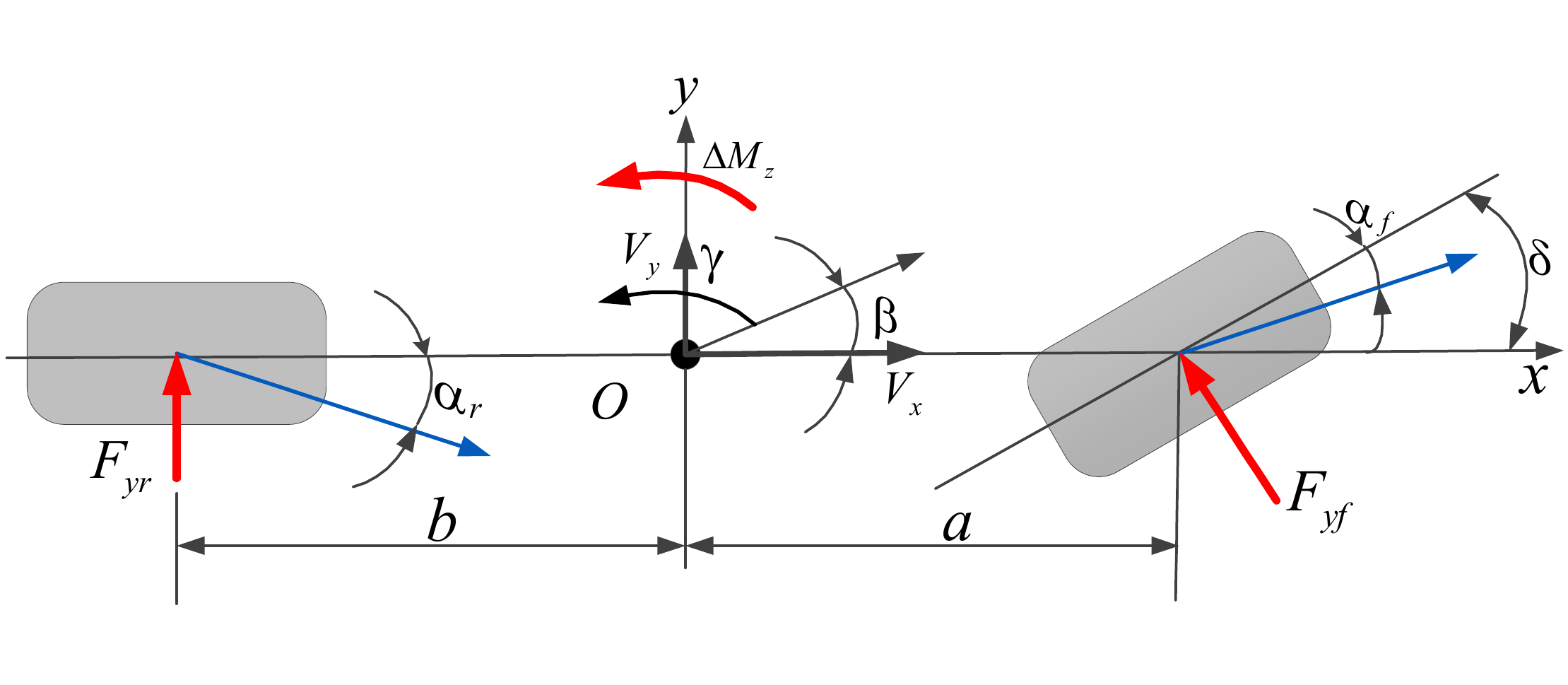}
	\caption{Single track vehicle model, $\alpha_f$ and $\alpha_r$ denote the front and rear tire slip angle, for detailed explanation please refer to \cite{zhang2018cooperative}\cite{huang2019novel}}
	\label{Fig1}
\end{figure}

The tire lateral slip rate, i.e. slip angle as shown in Fig. \ref{Fig1}, is a very important parameter used in estimating tire forces and vehicle sideslip angle \cite{hsu2009estimation}.~Its direct measurement or indirect estimation by intelligent tire technology can simplify the vehicle parameter estimation process by eliminating the sensors of steering angle, yaw rate, and even the observers. Yet it can improve the accuracy and reliability of vehicle states, leading to a better estimation of tire forces consequently.~Direct measurement by optical sensors is proposed in \cite{lamy2007comparison} but the conclusion is that they can not obtain the satisfactory performance as expected.~In \cite{erdogan2011tire},~the tire slip angle is estimated by correlating the lateral deformation of the inner liner of the tire.~It was shown that the slope of this deformation is proportional to the slip angle.~Such conclusion also is drawn from other studies \cite{matsuzaki2015intelligent} and \cite{erdogan2009novel}.~Although the estimated slip angle is proved to be able to improve vehicle stability and safety under the tested scenarios \cite{cheli2007methodology}, \cite{acosta2018virtual}, and \cite{erdogan2011tire},~the slip angle can only be accurately estimated up to 5 degrees,~which is not satisfactory in real-world driving cases.

As a result,~this study presents an estimation method for the tire slip angle based on the measured accelerations from the accelerometers, attached to the inner liner of the tire. Since almost all the related literature analyzes the time-domain measurements, this study attempts to find the correlation in the frequency domain.~Three different machine learning techniques are implemented to map the acceleration data to the tire slip angles. The phenomenon that the amplitude of the acceleration at around 1000Hz is clearly related to the tire slip angle is the first time revealed.~In addition, a comparative analysis is provided between time and frequency domains for tire slip angle estimations.~It is shown that the frequency domain analysis can estimate slip angles with higher accuracy comparing to machine learning algorithms trained by time domain data.~The obtained accuracy based on the frequency domain is precise enough for vehicle dynamics control to ensure the satisfactory control performance.

This paper is organized as follows: Section II presents the experimental system and the testing scenarios.~In Section III, the experimental data is analyzed in both time and frequency domain.~In Section IV, the results are analyzed and the neural network is designed to predict the slip angle. Section V presents a comprehensive comparison between different algorithms to show why the neural network based on Rprop algorithm is selected for this study.~Finally, the conclusion and future works are presented.	
	
\section{The Experimental System}	
	
This section briefly shows the intelligent tire system, the Measure Test Simulate (MTS)  tire testing apparatus, and their interaction for experiments. In addition, a wide range of experimental scenarios are defined to collect the required data.

\subsection{Intelligent Tire system based on MEMS Accelerometers}	
	
\begin{figure}[!htb]
	
	\subfloat[]{\includegraphics[width=0.5\linewidth,height=0.5\linewidth]{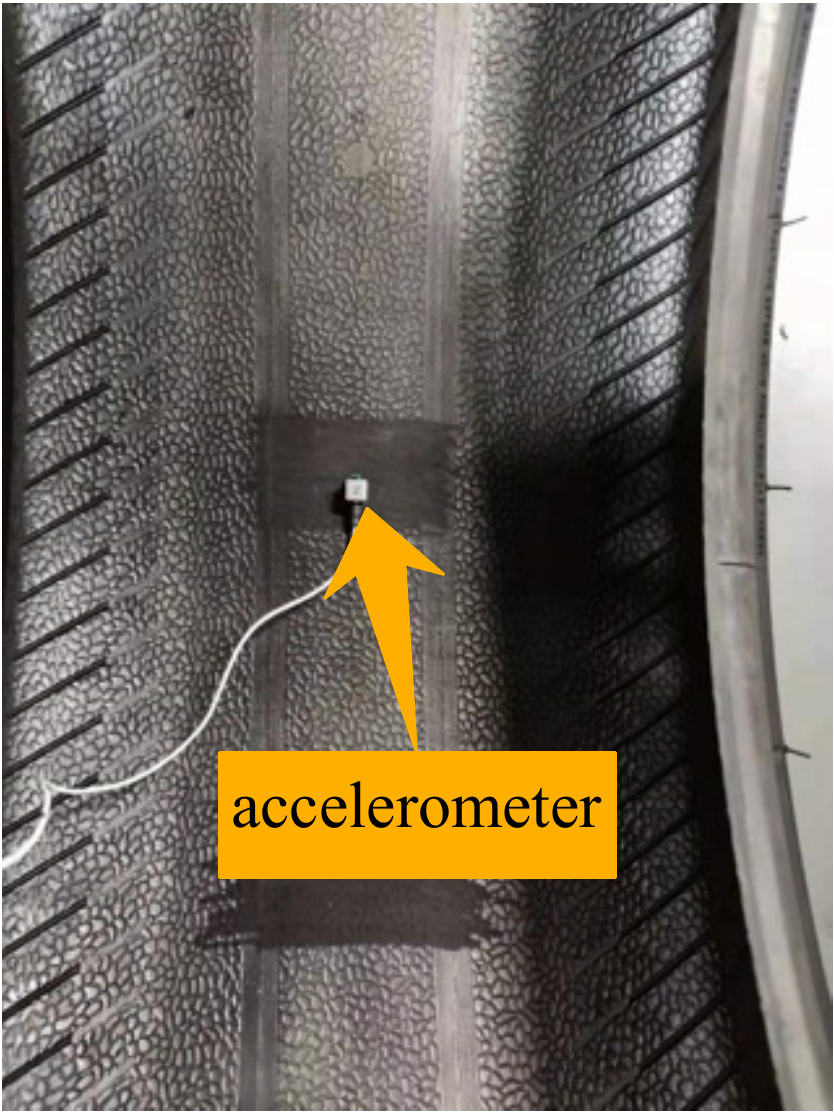}}
	\subfloat[]{\includegraphics[width=0.5\linewidth,height=0.5\linewidth]{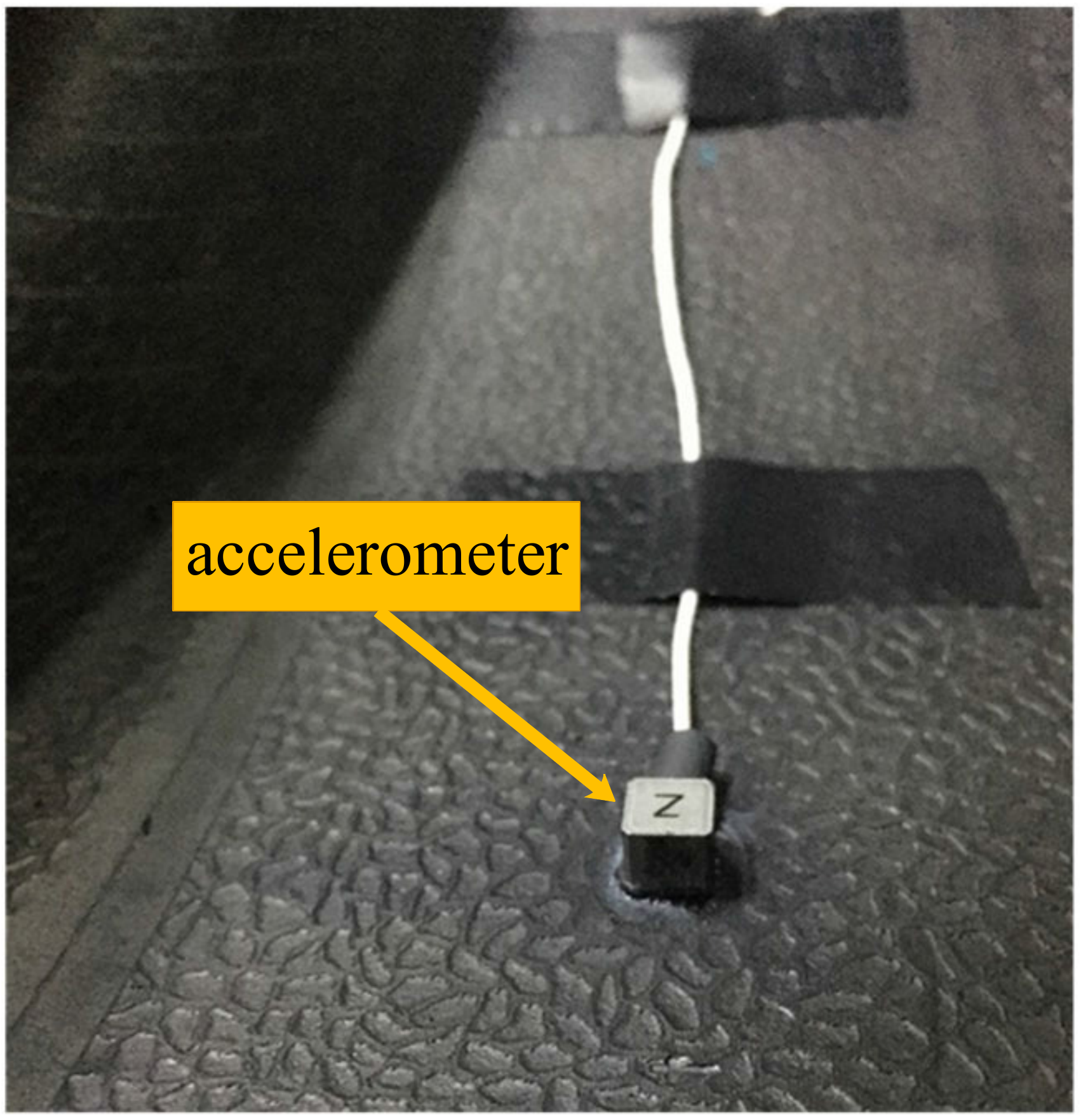}}
	\caption{Tri-axis accelerometer attached to the inner liner of the tire}
	\label{Fig2}
	\vspace{-0.5em}
\end{figure}

The intelligent tire testing system consists of a tri-axial acceleration sensor, a slip ring, a signal regulator, and a  National Instrument (NI) data acquisition system (DAQ). The tri-axial acceleration sensor is glued and fixed at an arbitrary point along the inner liner of the tire as shown in Fig.\ref{Fig2}, and used to measure the accelerations in longitudinal, lateral, and vertical directions. 

\begin{figure}[!htb]
		\centering
		\vspace{-1em}
		\includegraphics[width=0.5\linewidth]{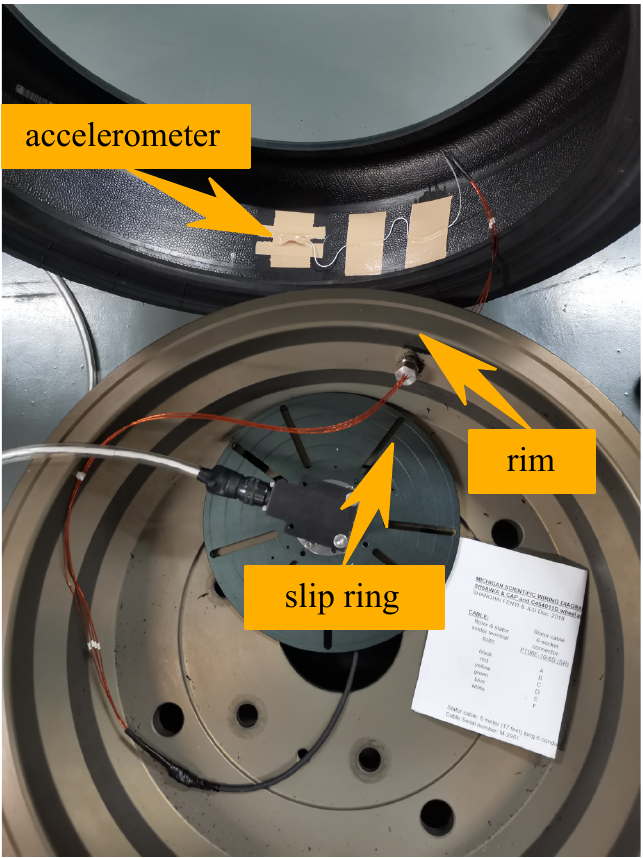}
		\caption{Intelligent tire system}
		\label{Fig3}
	\end{figure}

A slip-ring device, shown in Fig.\ref{Fig3}, is mounted to the rim, which transmits the sensor signals from the rotating tire to the NI DAQ system through the signal regulator.~The signal debugger provides energy supply for the signal,~while the NI acquisition system can adjust signal channel and sampling frequency to collect the acceleration signals. The sampling rate is 10kHz, which is enough for the purpose of this study.

\subsection{MTS tire testing system}
\begin{figure}[!htb]
	\centering
	\vspace{-1em}
	\includegraphics[width=0.8\linewidth]{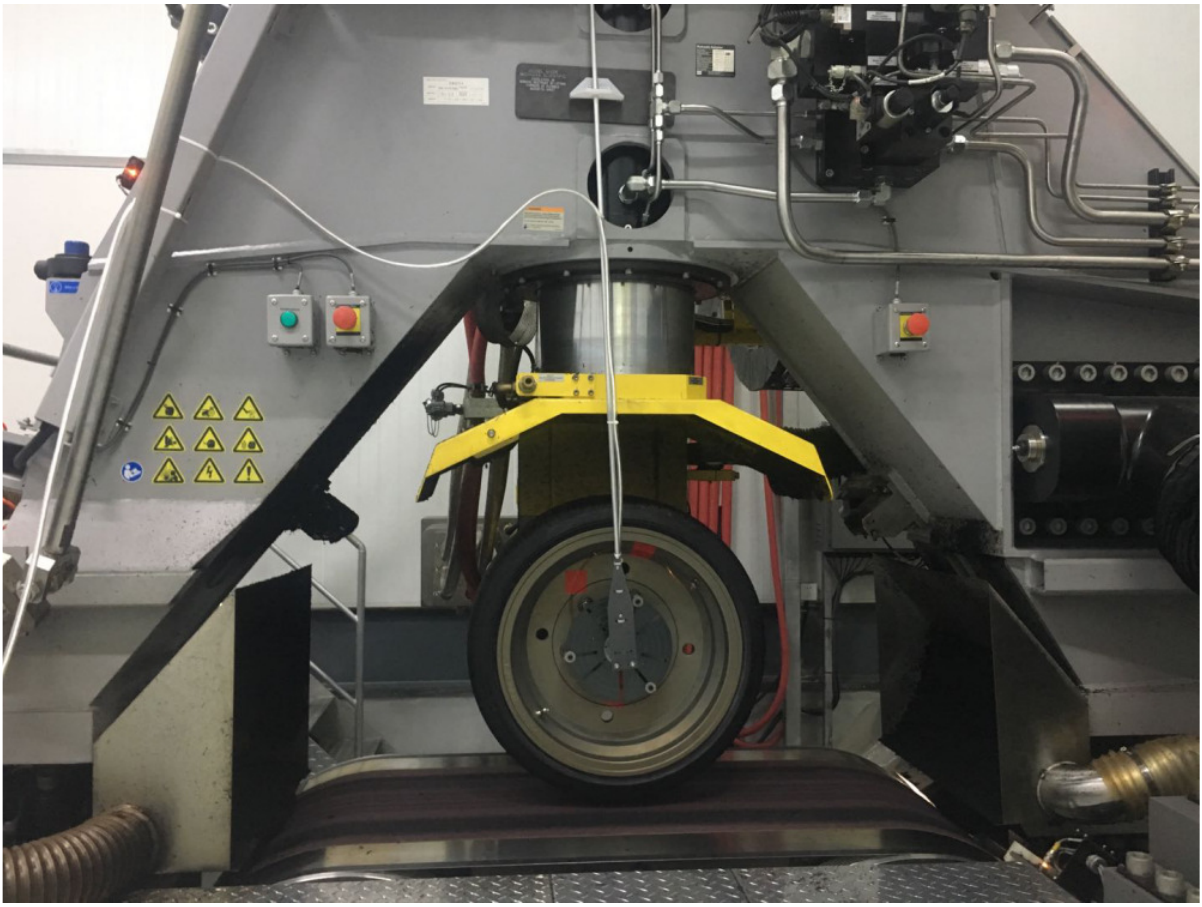}
	\caption{MTS flat-trac tire testing platform with the intelligent tire system}
	\label{Fig4}
\end{figure}

\begin{figure}[!htb]
	\centering
	\vspace{-1em}
	\includegraphics[width=\linewidth]{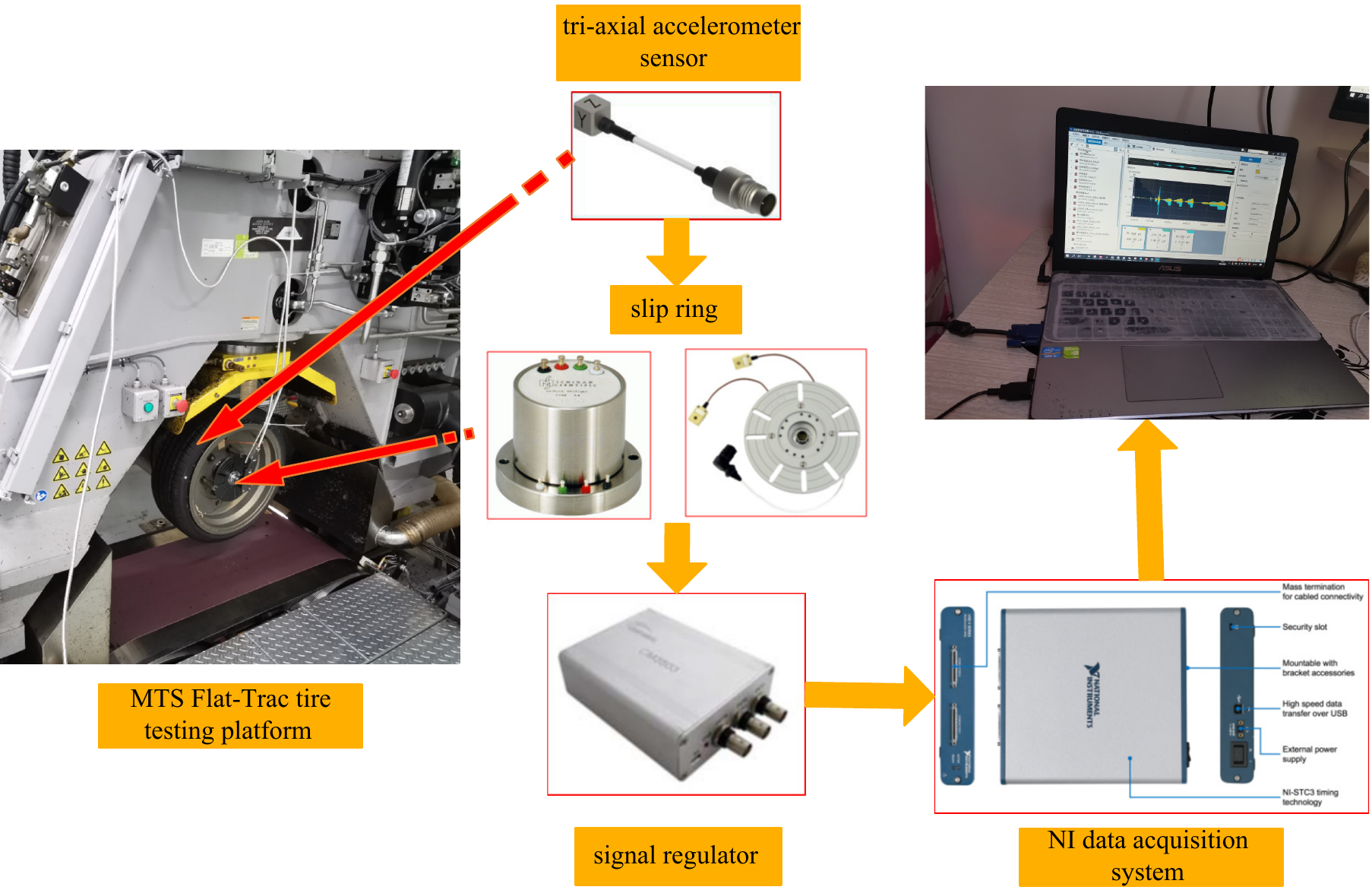}
	\caption{The entire testing system}
	\label{Fig5}
\end{figure}

To extract the acceleration data for different tire slip angles,  the MTS Flat-Trac testing system is used as shown in Fig.\ref{Fig4}. The required vertical load, driving torque, speed, and tire slip angle can be applied to simulate tire working under different conditions. At the same time, the acceleration signals are recorded by the NI DAQ system. In addition, the slip angles and tire forces in six directions can be measured and recorded by the DAQ system of the MTS test-bed. The diagram showing the whole testing system is presented in Fig.\ref{Fig5}.

\subsection{Experimental scenarios}
In this section, the experimental scenarios are designed and shown in Table \ref{Table1}.~In order to develop the tire slip angle estimation methods, a wide range of cornering maneuvers are conducted with different vertical loads, velocity, and slip angles. Since machine-learning based methods are used in this study, the data should be rich enough to train the algorithms. In total, the data consists of 4705 samples (3552 step, 1153 triangular), and each of them is divided into training (70\%) and testing data sets (30\%).

\begin{table}[!htb]
	\caption{Testing scenarios}\label{Table1}
	\begin{tabularx}{\linewidth}{cX<{\centering}}
		\toprule
		Cornering & Parameters \\
		\midrule
		Pressure (kPa) & 220  \\
		& \\
		Vertical Load (N) & 2080,4160,6240\\
		&\\
		Velocity (km/h) & 30,60  \\
		& \\
		Slip Angle (deg) & $\pm8,\pm7,\pm6,\pm5,\pm4, \pm3,\pm2,\pm1$,Triangular Wave (up to 10 degrees) \\
		\bottomrule
	\end{tabularx}
\end{table}

\section{Experimental data analysis}
In this section, the data collected from the aforementioned experimental system is demonstrated in both time and frequency domain and the related results are also presented. 

\begin{figure}[!htb]
	\centering
	\subfloat[]{\includegraphics[width=0.5\linewidth,height=0.3\linewidth]{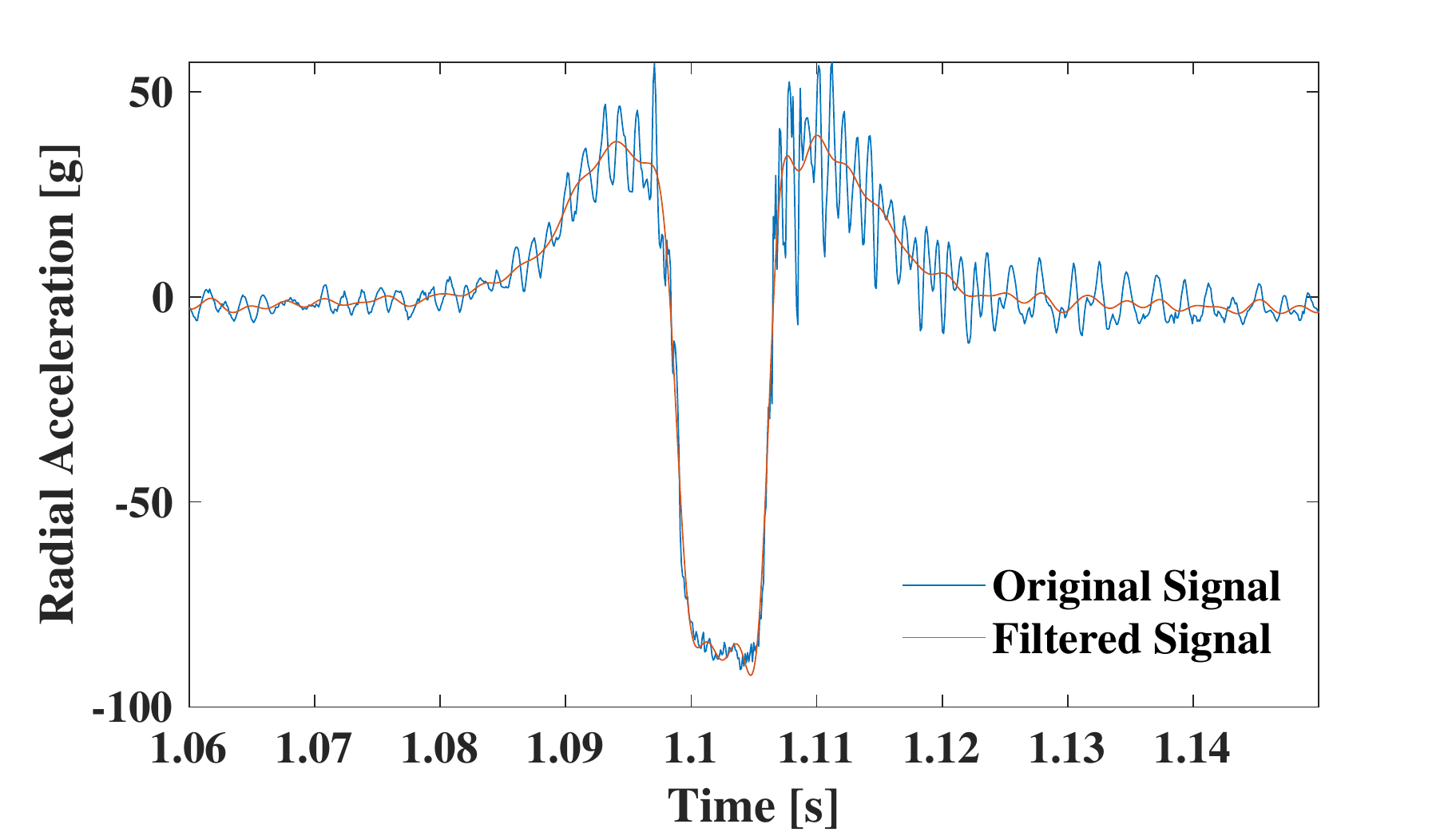}}
	\subfloat[]{\includegraphics[width=0.5\linewidth,height=0.3\linewidth]{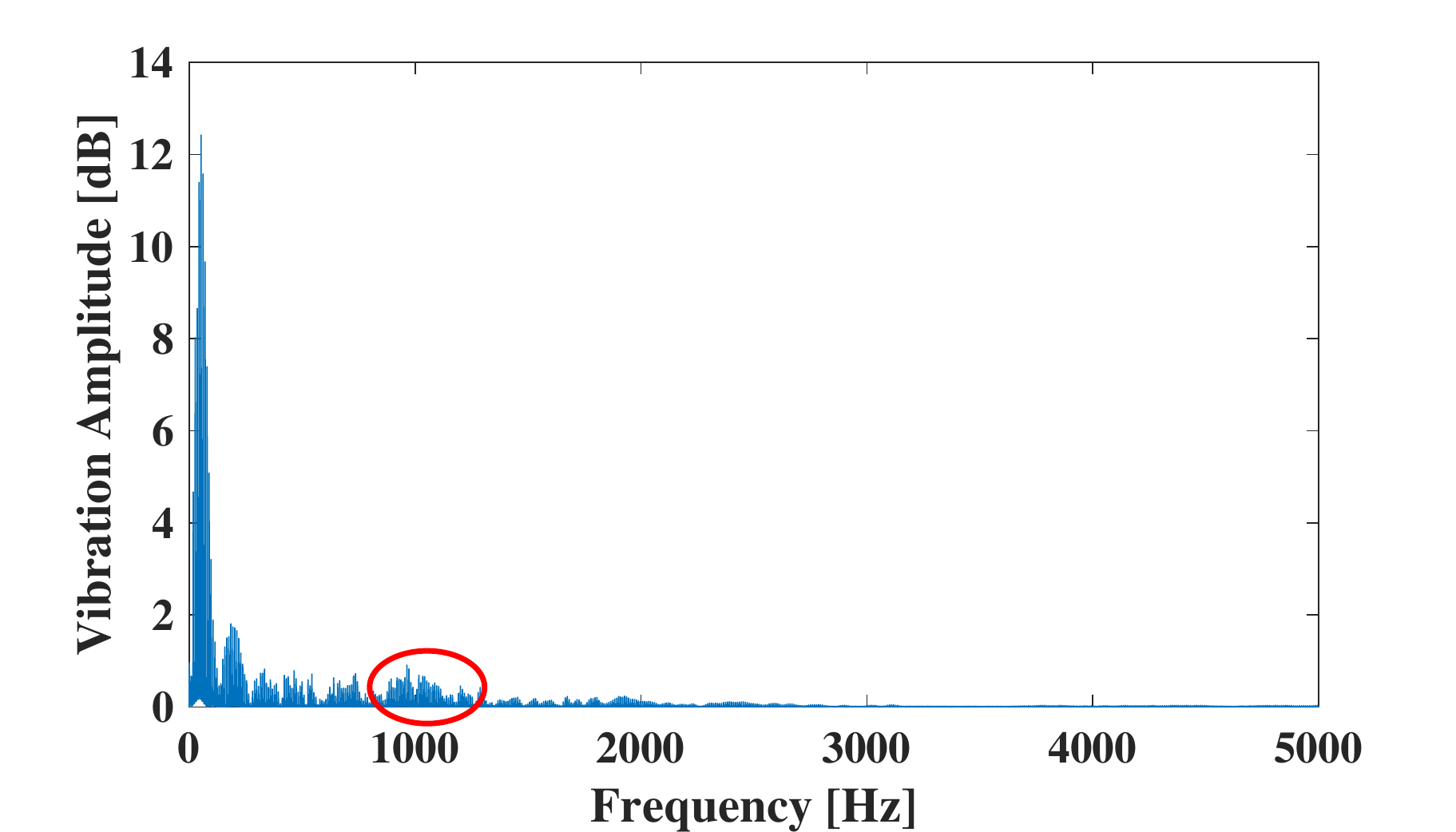}}
	\caption{Acceleration in both (a) time and (b) frequency domains under 0 deg slip-angle condition (60km/h at 6240N load)}
	\label{Fig6}
	\vspace{-0.5em}
\end{figure}

\begin{figure}[!htb]
	\centering
	\subfloat[]{\includegraphics[width=0.5\linewidth,height=0.3\linewidth]{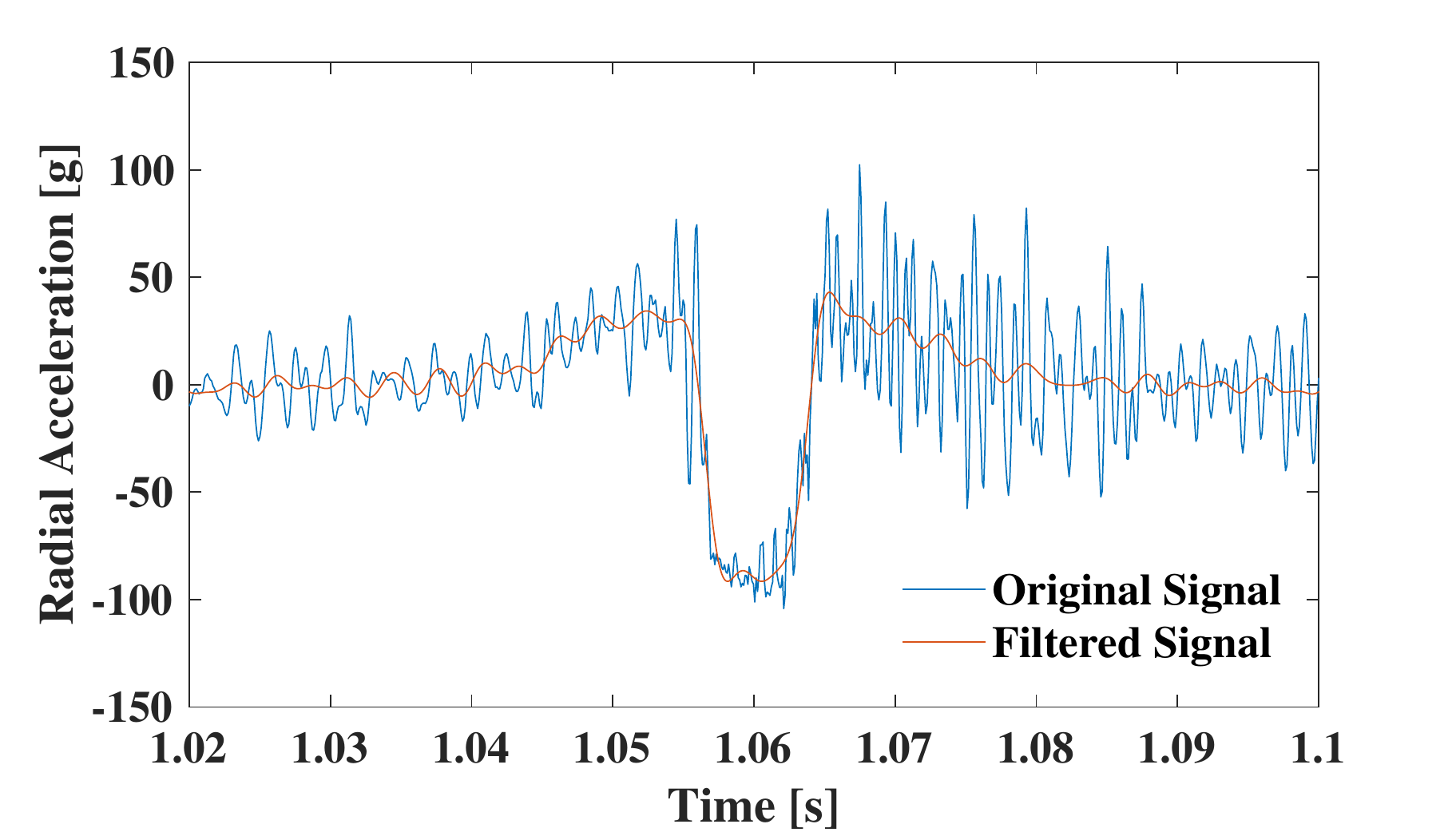}}
	\subfloat[]{\includegraphics[width=0.5\linewidth,height=0.3\linewidth]{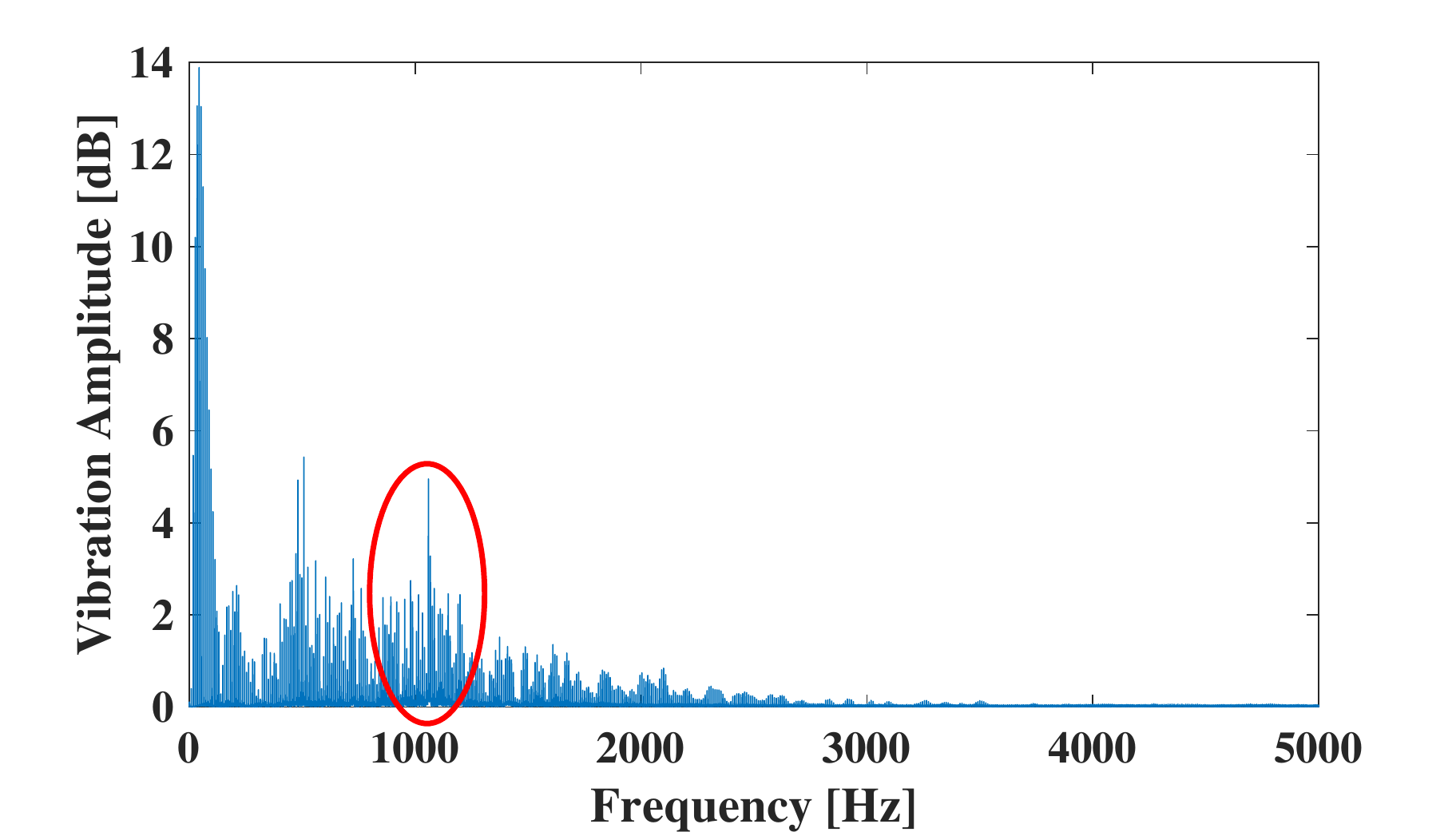}}
	\caption{Acceleration in both (a) time and (b) frequency domains under 2 deg slip-angle condition (60km/h at 6240N load)}
	\label{Fig7}
	\vspace{-0.5em}
\end{figure}

\begin{figure}[!htb]
	\centering
	\subfloat[]{\includegraphics[width=0.5\linewidth,height=0.3\linewidth]{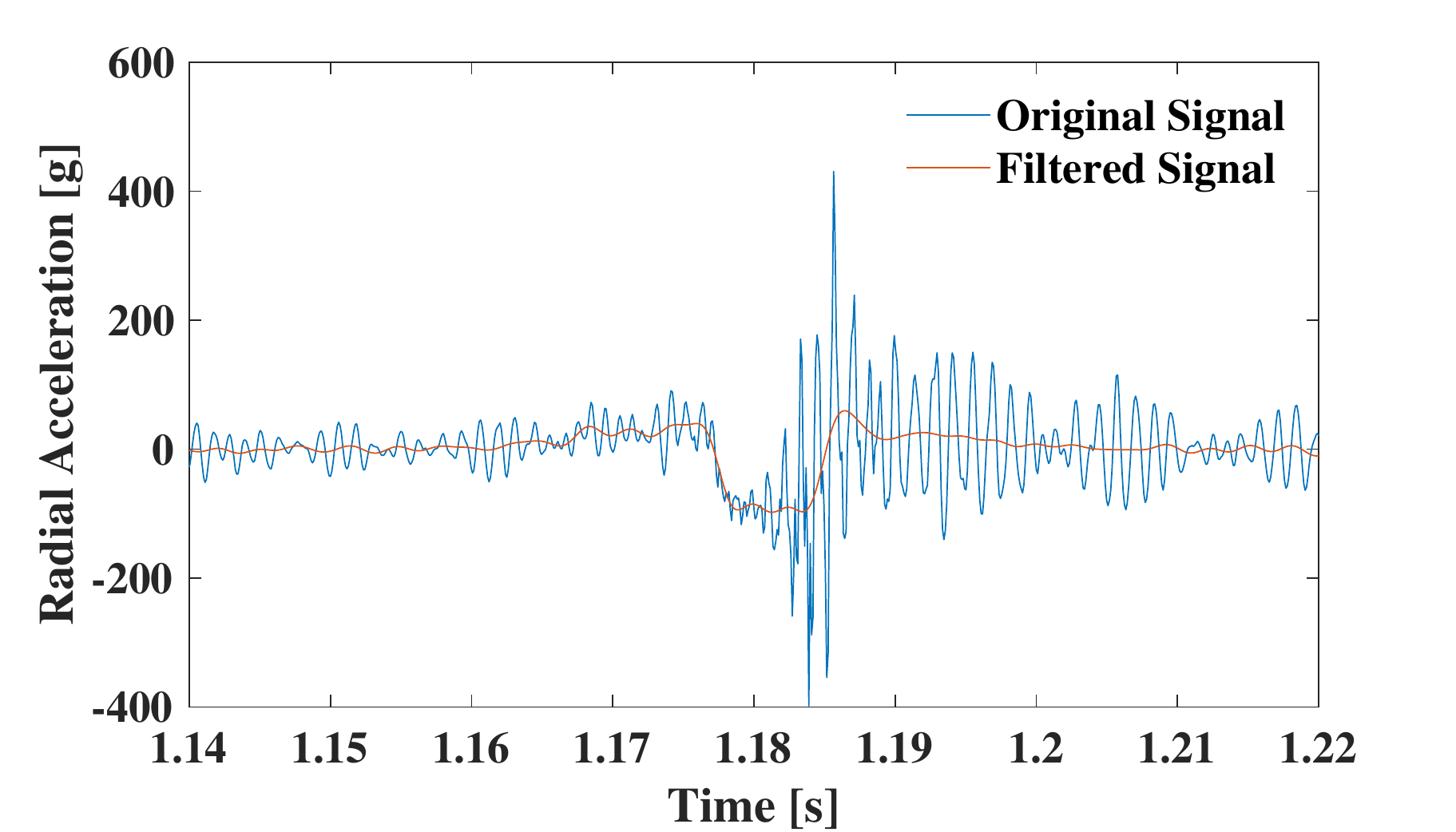}}
	\subfloat[]{\includegraphics[width=0.5\linewidth,height=0.3\linewidth]{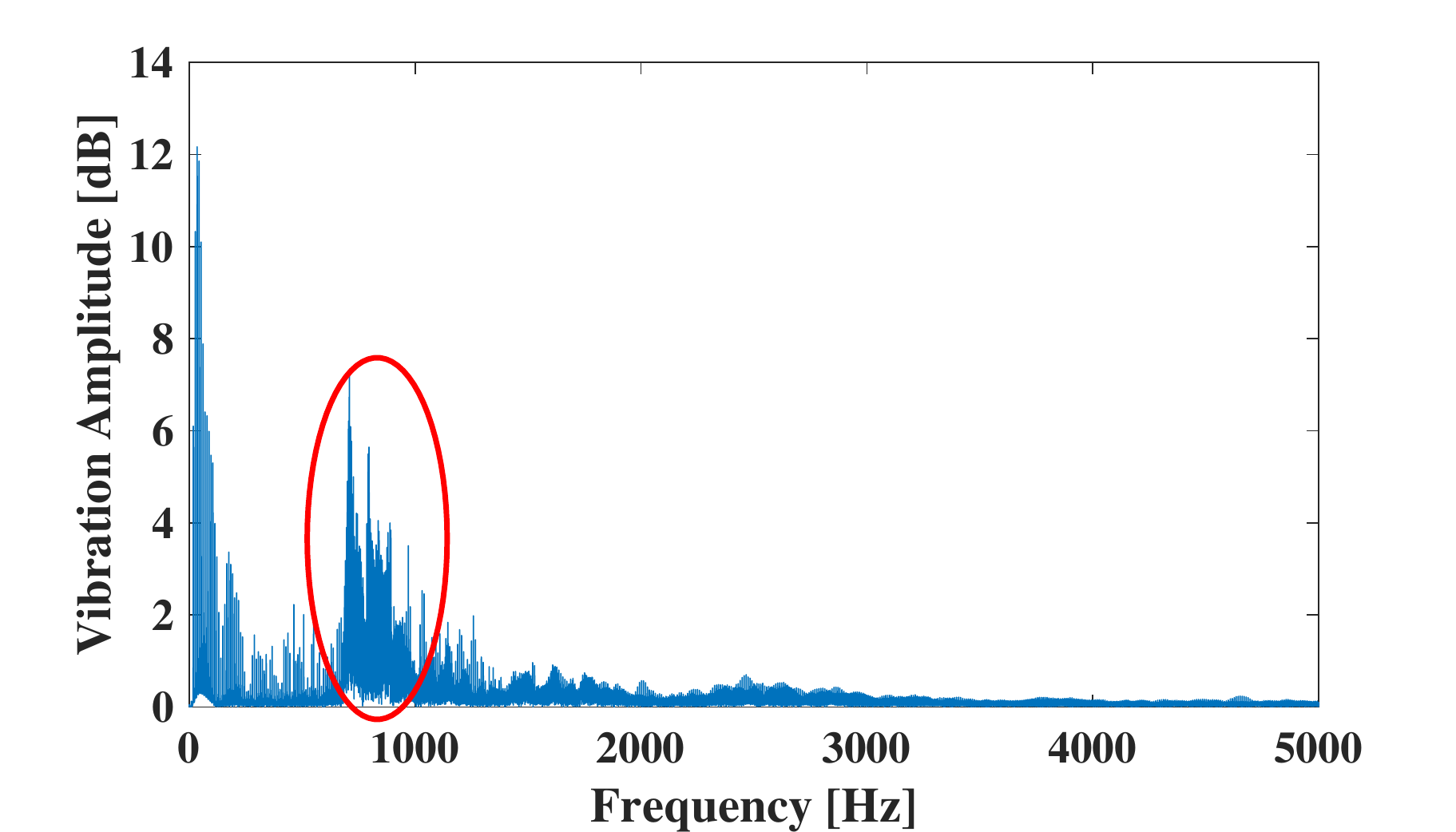}}
	\caption{Acceleration in both (a) time and (b) frequency domains under 4 deg slip-angle condition (60km/h at 6240N load)}
	\label{Fig8}
	\vspace{-0.5em}
\end{figure}

\begin{figure}[!htb]
	\centering
	\subfloat[]{\includegraphics[width=0.5\linewidth,height=0.3\linewidth]{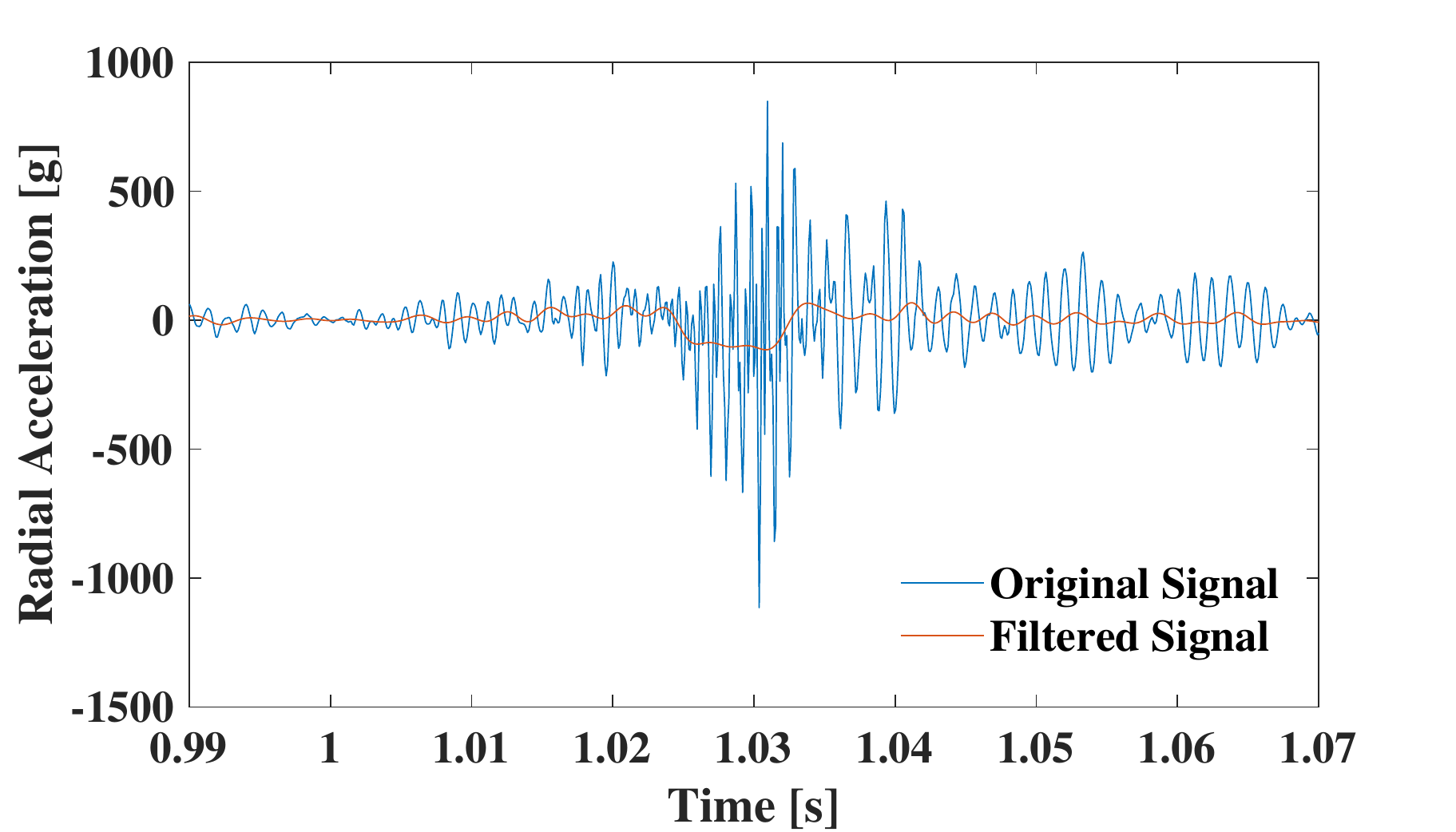}}
	\subfloat[]{\includegraphics[width=0.5\linewidth,height=0.3\linewidth]{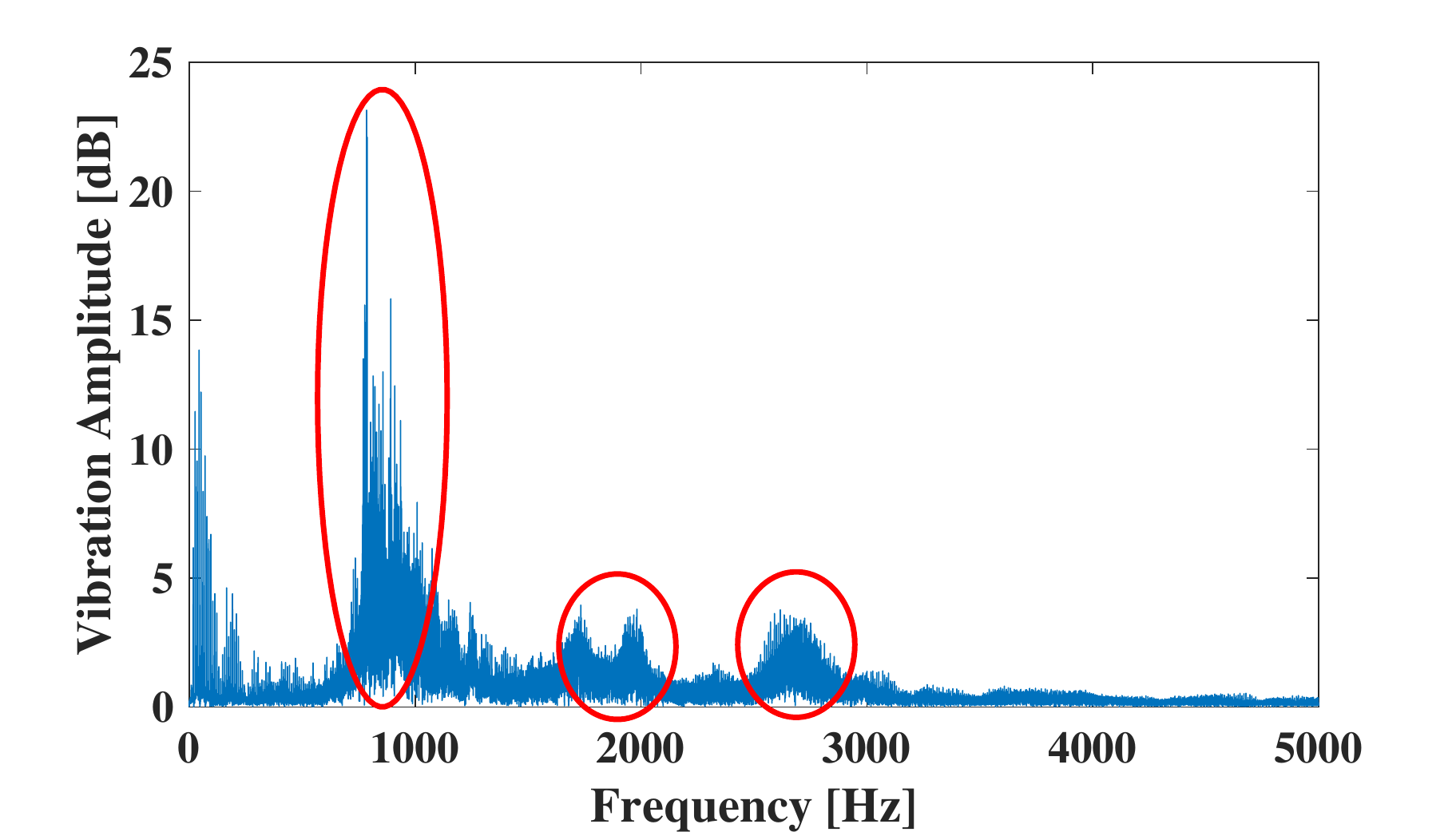}}
	\caption{Acceleration in both (a) time and (b) frequency domains under 6 deg slip-angle condition (60km/h at 6240N load)}
	\label{Fig9}
	\vspace{-0.5em}
\end{figure}

The vertical acceleration responses  with different slip angles in both time and frequency domain are presented. As stated in the previous section, the sampling frequency of the raw data is 10kHz. After applying the FFT, the frequency characteristics of the test data under different tire slip-angle conditions are presented  in Fig. \ref{Fig6} to Fig. \ref{Fig9}. In addition, the red lines in Fig. 6(a) to Fig. 9(a) means the signal processed by a filter with a 400Hz cutoff frequency.

$Remarks$: in this study, although the vertical acceleration is used, the lateral or circumferential ones exhibit the same characteristics and the similar conclusions can be drawn as well. 

\begin{figure}[!htb]
	\centering
	\vspace{-1em}
	\includegraphics[width=\linewidth]{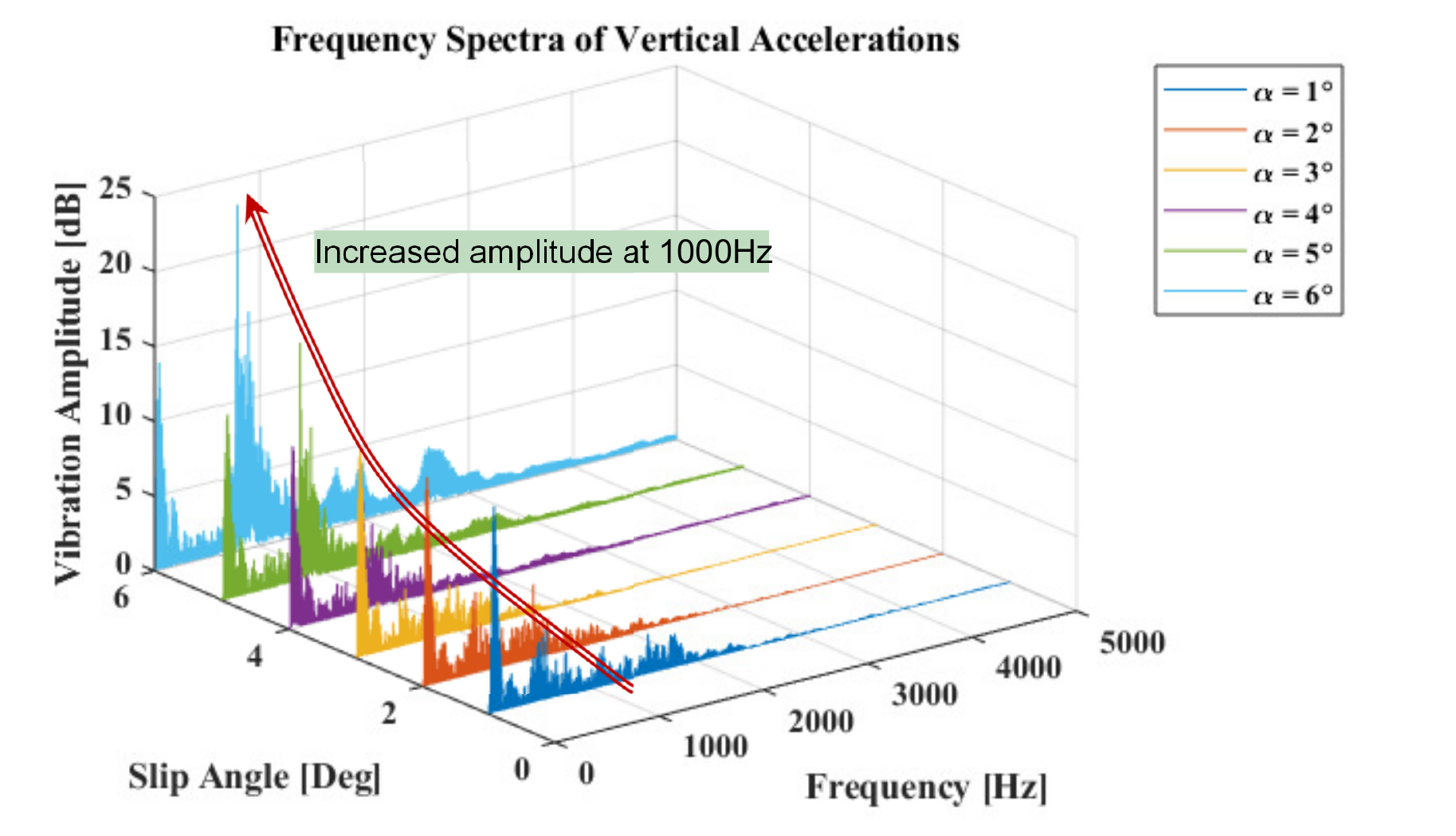}
	\caption{Frequency spectrum of vertical accelerations w.r.t. different silp angles}
	\label{Figure10}
\end{figure}

Based on the results shown in Fig. 6(a) to Fig. 9(a), it is difficult to conclude any obvious correlation between the tire slip angles and the acceleration values.~However, from Fig. 6(b) to Fig. 9(b) an interesting phenomenon can be observed that the amplitudes around 1000Hz directly relate to the slip angles. More specifically, with the increase of the slip angle, the amplitude values tend to be larger and an obvious spike appears. This interesting result is even more pronounced  in Fig. \ref{Figure10}. When the slip angle is larger than 6 degrees, spikes also happen at around 2000Hz and 3000Hz, respectively. 

In addition, the high-frequency oscillations can be seen in Fig. 6(a) to Fig. 9(a), especially at the rear part of the contact patch, which will be further explained in Section IV. Such oscillations extend to the front part of the contact patch as the slip angle becomes larger. The reason is that in the slip region of the contact patch, the micro-vibration can be easily generated, resulting into a piercing noise. The accelerometer captures the noises or micro-vibrations in terms of the high-frequency oscillations. Furthermore, as the slip angle increases, the slip region extends as well until it spreads to the whole contact patch.  

\section{Tire slip angle estimation algorithm}

In this section, the estimation algorithm based on a neural network is elaborated and the identification results are presented by comparing to the time- and frequency-domain estimation methods. The performance of the estimation accuracy is quantified by the normalized root mean square error (NRMS).

\subsection{Data processing}

In our data processing, we have used the acceleration data  obtained in the contact patch region and the acceleration data related to the other places of tire have been neglected.~In data preprocessing stage, the raw data needs to be processed and then used for the machine-learning algorithm, which is crucial for the neural network training.~The whole process for time-domain data includes filtering, useful information extraction, data transformation, and data normalization.~Meanwhile, Fast Fourier Transform (FFT) is exploited to process the raw data for frequency-domain data analysis. The whole process is briefly introduced in this section and also shown in Fig. \ref{Fig11}: 

\begin{figure}[!htb]
	\centering
	\vspace{-1em}
	\includegraphics[width=\linewidth]{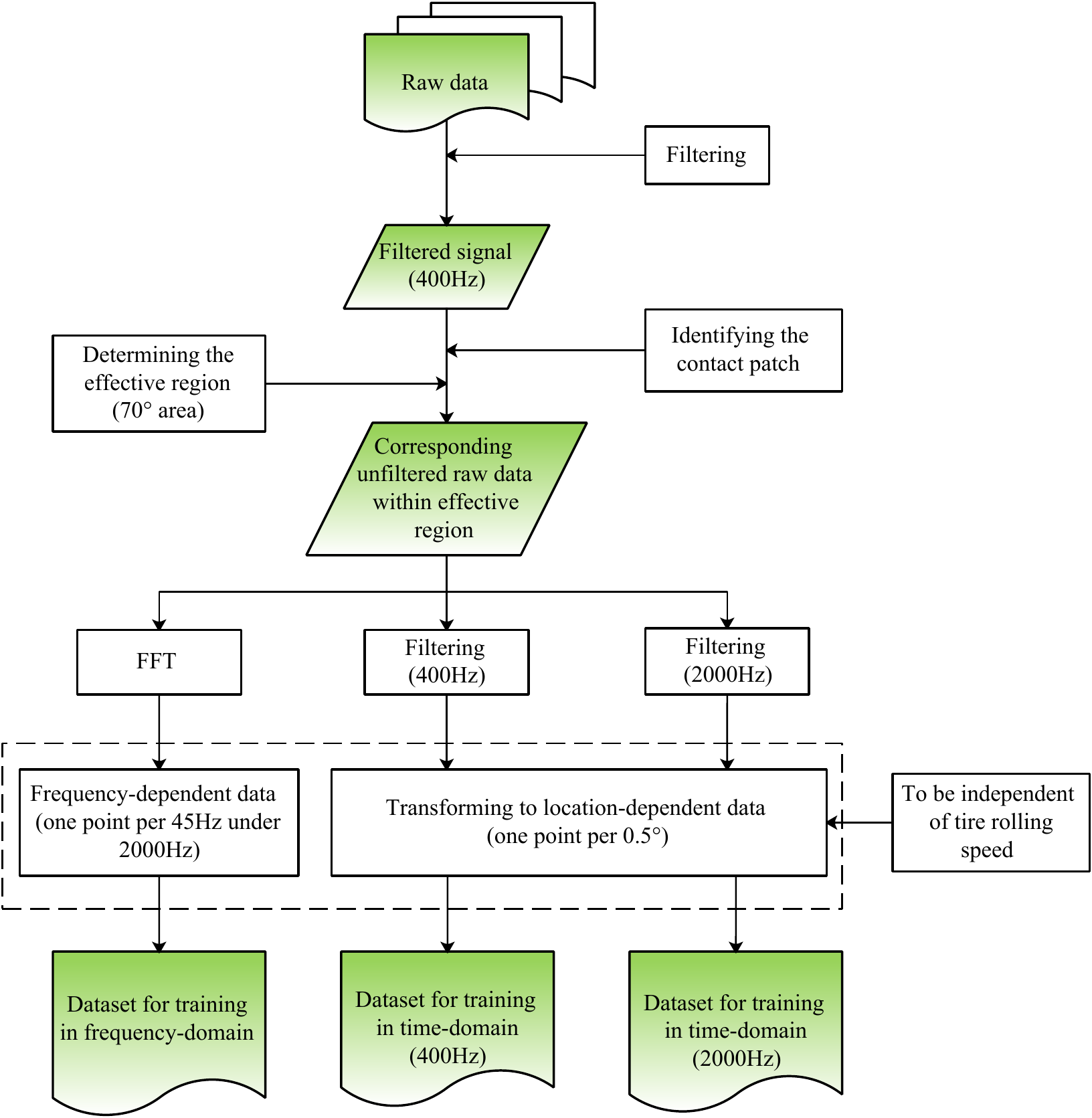}
	\caption{Flowchart of data process}
	\label{Fig11}
\end{figure}

\begin{enumerate}[1)]
	\item $Filtering$: the sampling frequency of the collected acceleration signal is 10kHz, which is processed  by a 5 order Butterworth low-pass filter with a tuned cut-off frequency (i.e. 400Hz) to obtain the contact patch.~However, the filtered data in higher frequency will be used for road condition identification in future studies.
	\item $Identifying$ $the$ $contact$ $patch$: two peaks can be observed obviously in the circumferential direction when the accelerometer enters and leaves the contact patch. Therefore, the data for every tire revolution can be extracted by using the encoder signal such that the acceleration peaks, which represents the contact patch will be then detected.~As shown in Fig. \ref{Fig12}, the points B and D can be adopted as the starting and ending position of the contact patch, respectively, under the test condition.~The same method applies to other conditions as well. 
	\item $Independency$ $from$ $rotating$ $speed$: to simplify the structure of the neural network and make the identification method independent of the tire rolling speed, the time-dependent measurement is processed as presented in Fig. \ref{Fig11}. Specifically, by using the encoder signal and the current tire rolling speed, the acceleration for any point along the inner liner can be obtained.~After identifying the contact patch in step 2, the center of contact patch C can be determined as shown in Fig. \ref{Fig12}.~For any testing conditions, the angle of the contact patch is less than $ {70}^{\circ}$.~Then, from the center point C, point A and E are found with a whole angle of ${70}^{\circ}$.~Finally, one single data per ${0.5}^{\circ}$ within point A and E is extracted as the inputs to train the neural network.~In this way, no matter under which working condition, the used acceleration data is always from the same locations. It should be noted that, for the sake of comparison, the data is filtered by two frequencies i.e. 400 Hz and 2000 Hz. 
	
	\begin{figure}[!htb]
		\centering
		\vspace{-1em}
		\includegraphics[width=0.8\linewidth]{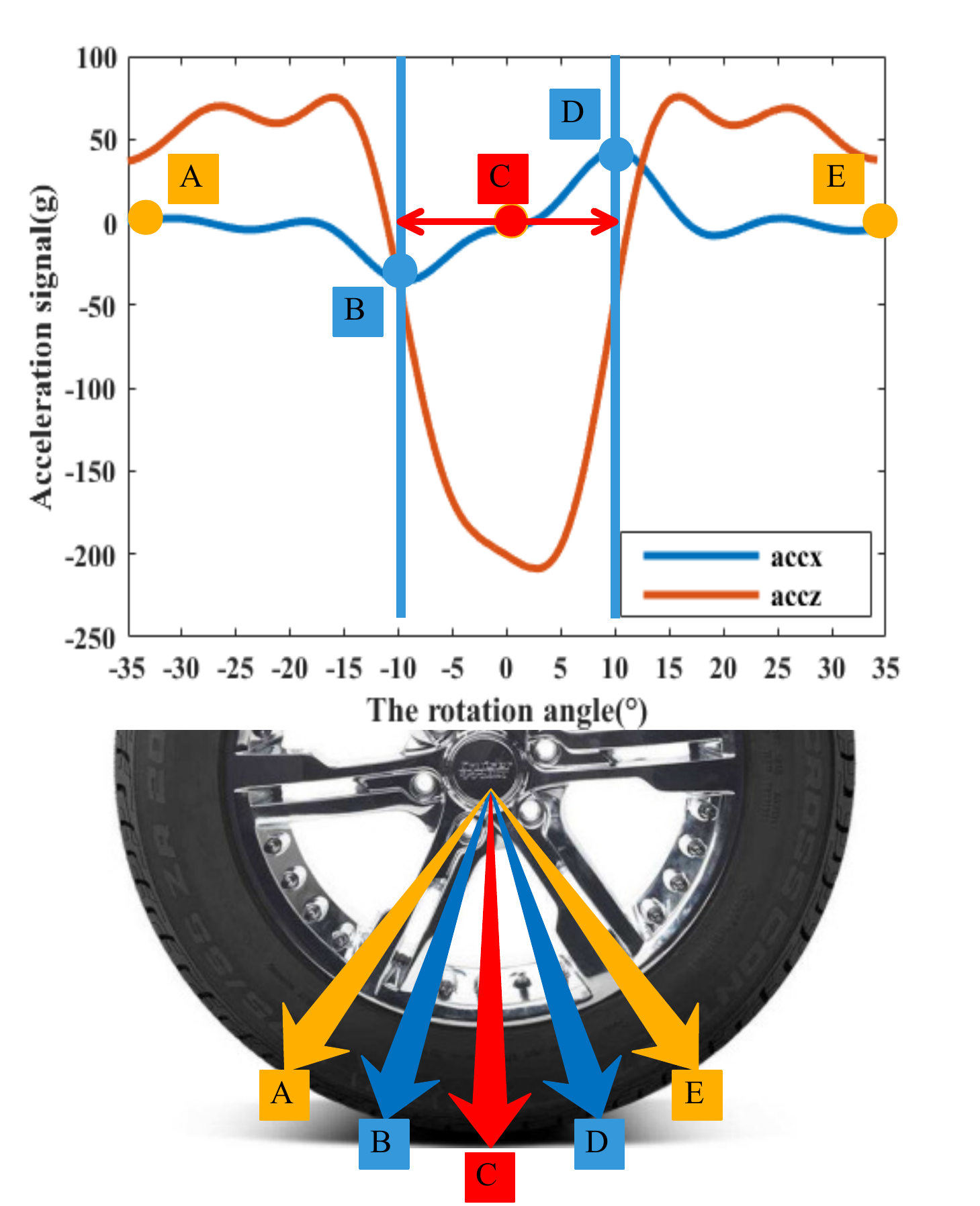}
		\caption{Contact patch and the featured region extraction}
		\label{Fig12}
	\end{figure}

    \item $FFT$: the raw data is processed by the FFT, as mentioned above, the featured region is around 1000Hz such that the data under 2000Hz is used and the interval is 45 Hz between each point. The inputs to train the network are the amplitudes of 40 points while the output is one single scalar, the value of slip angle. 
	
	\item $Normalization$: the Min-Max normalization is used to facilitate the neural network training in this study. The min-max normalization is a linear transformation and it can preserve all relationships of the data values exactly. It is represented as below:
	\begin{align}
	{x_{norm}} &= \frac{x-x_{min}}{{x_{max}-x_{min}}}
	\end{align}
	where $x$ are the measured acceleration values and $x_{min}$ and $x_{max}$ are the minimum and maximum of the collected data. 
	\\
	
	
\end{enumerate}	

\subsection{Neural network training}	

An artificial neural network (ANN) based on Rprop algorithm  is developed to estimate the tire slip angles by using the measured tire acceleration at the attached point. Several algorithms have been developed for neural networks learning. For example, Broyden-Fletcher-Goldfarb-Shanno (BFGS), Levenberg-Marquardt, and conjugate gradients are classified as the well-known algorithms for training feedforward neural networks. Gradient descent methods (GDM) are also popular for supervised learning of neural networks. One of the most efficient techniques based on GDM is batch back-propagation, which minimizes the error function implementing steepest descent method.~Adaptive gradient-based algorithms have been also exploited for training neural networks.~They are considered as one of the trendiest algorithms for optimization, and also machine learning.~In this study, the Rprop algorithm is considered as the main method because it is recognized as one of the most used approaches as per the convergence speed, accuracy, and the robustness to the learning parameters and rate. The Rprop method adopts a sign-based technique and updates the weights to prevent detrimental influence of derivatives' magnitude on the updated weights. Since Rprop is able to effectively tackle the noisy error, it is appropriate for application in hardware \cite{askari2019towards}.

\begin{figure}[!htb]
	\centering
	\vspace{-1em}
	\includegraphics[width=\linewidth]{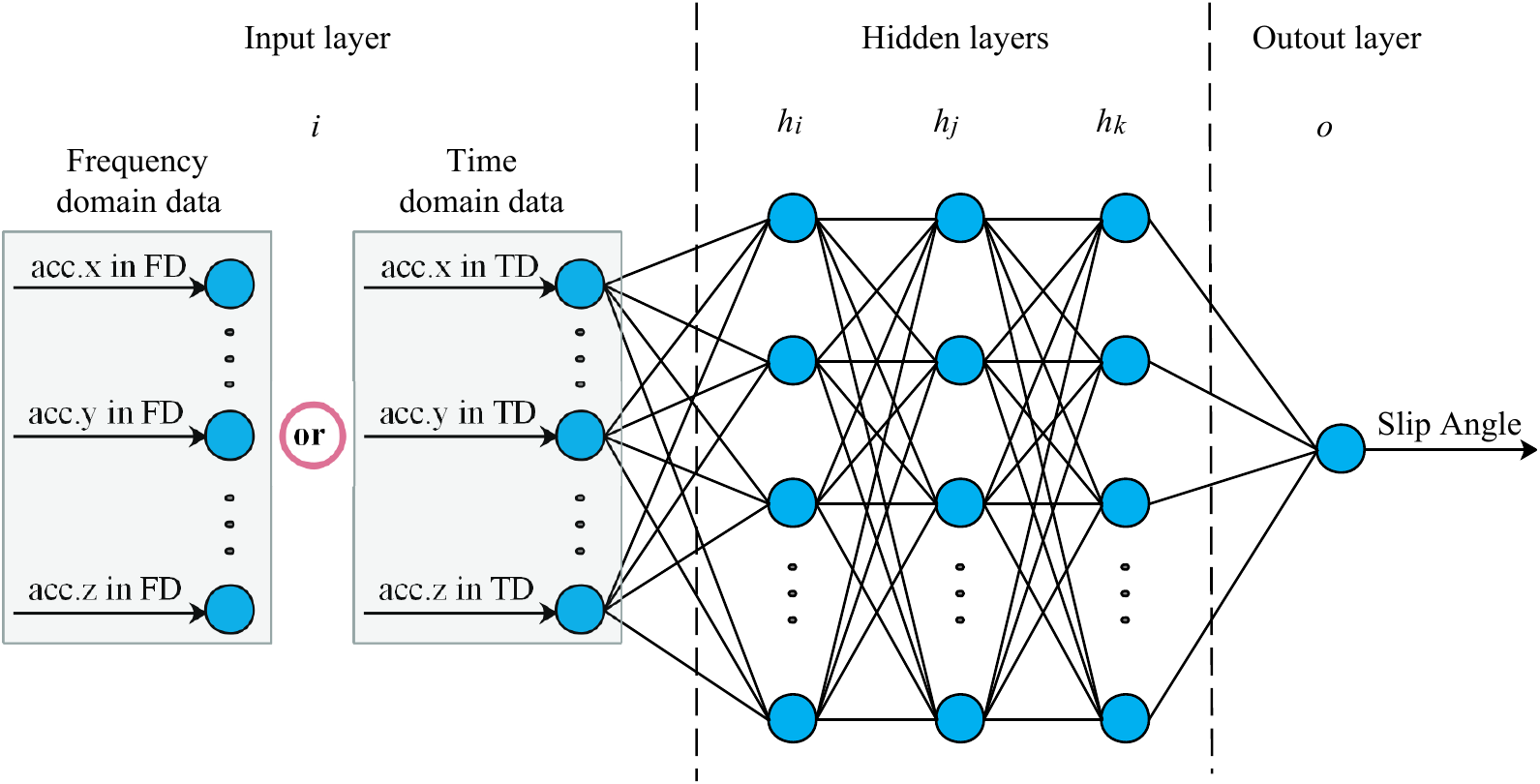}
	\caption{The structure of the designed neural network}
	\label{Fig13}
\end{figure}

The structure of the neural network ,based on Rprop algorithm, used in this paper is shown in Fig. \ref{Fig13}, where three hidden layers are used in the designed neural network.~The logistic activation function and mean squared error (MSE) performance index are adopted with Resilient backpropagation (Rprop) algorithm. All the accelerations in three directions are used as the inputs, which contain 40 amplitude variables in each direction for the frequency domain. For the sake of comparison, we also use 140 variables in each direction for time-domain training, which corresponds to spatial points within the 70 angle range to cover contact patch. 

\begin{figure}[!htb]
	\centering
	\includegraphics[width=0.8\linewidth]{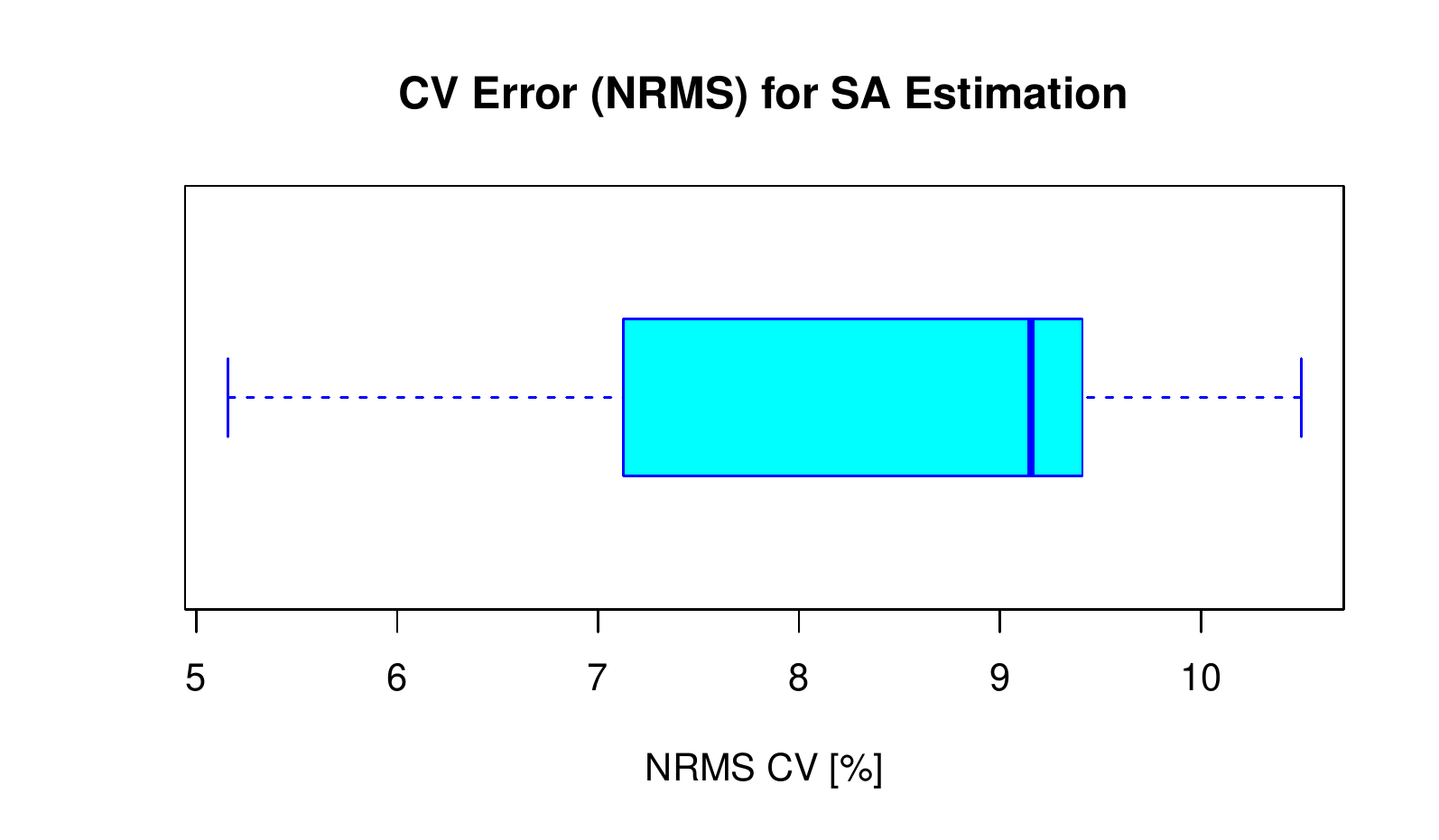}
	\caption{Boxplot of the 10-fold cross-validation results}
	\label{Fig14}
\end{figure}

Regarding the training work, all the collected data is divided into training and test sets. Training set is used to find the relationship between dependent and independent variables while the test set assesses the performance of the model. Cross-validation (CV) is a technique to assess the generalizability of a model to unseen data. A 10-fold cross-validation method is used to minimize the bias associated with the random sampling of the training.~In the 10-fold cross-validation, the data set is divided into 10 parts, and then 9 parts are used for training and 1 set is used for testing. The process was then repeated until all parts are tested. Table \ref{Table2} shows the fold-wise performance analysis and Fig. \ref{Fig14} shows a box plot of the cross-validation results. NRMS errors ranged from 5.16 percent to 10.49 percent for slip angle estimation, and the average NRMS error is 8.39 percent. The CV results show that the trained model is reliable and has an accurate estimation for tire slip angle even under extreme conditions. 

\begin{table}[htbp]
	\centering
	\caption{Fold-wise performance analysis}\label{Table2}
	\begin{tabular}{cccc}
		\toprule
		Folds & Training instances & Testing instances & NRMS errors(\%)  \\
		\midrule
		Fold 1 & 4220 & 485 & 9.39   \\
		Fold 2 & 4214 & 491 & 7.13   \\
		Fold 3 & 4228 & 477 & 9.20   \\
		Fold 4 & 4228 & 477 & 10.49  \\
        Fold 5 & 4234 & 471 & 10.47  \\
        Fold 6 & 4246 & 459 & 5.16   \\
        Fold 7 & 4230 & 475 & 5.96   \\
        Fold 8 & 4232 & 473 & 9.41   \\
        Fold 9 & 4266 & 439 & 9.11   \\
        Fold 10 & 4247 & 458 & 7.62  \\
        \midrule
        \multicolumn{3}{c}{Average error} & 8.39        \\
		\bottomrule
	\end{tabular}
\end{table}

\subsection{Neural network prediction and validation for frequency data}

\begin{figure}[!htb]
	\centering
	\vspace{-1em}
	\includegraphics[width=\linewidth]{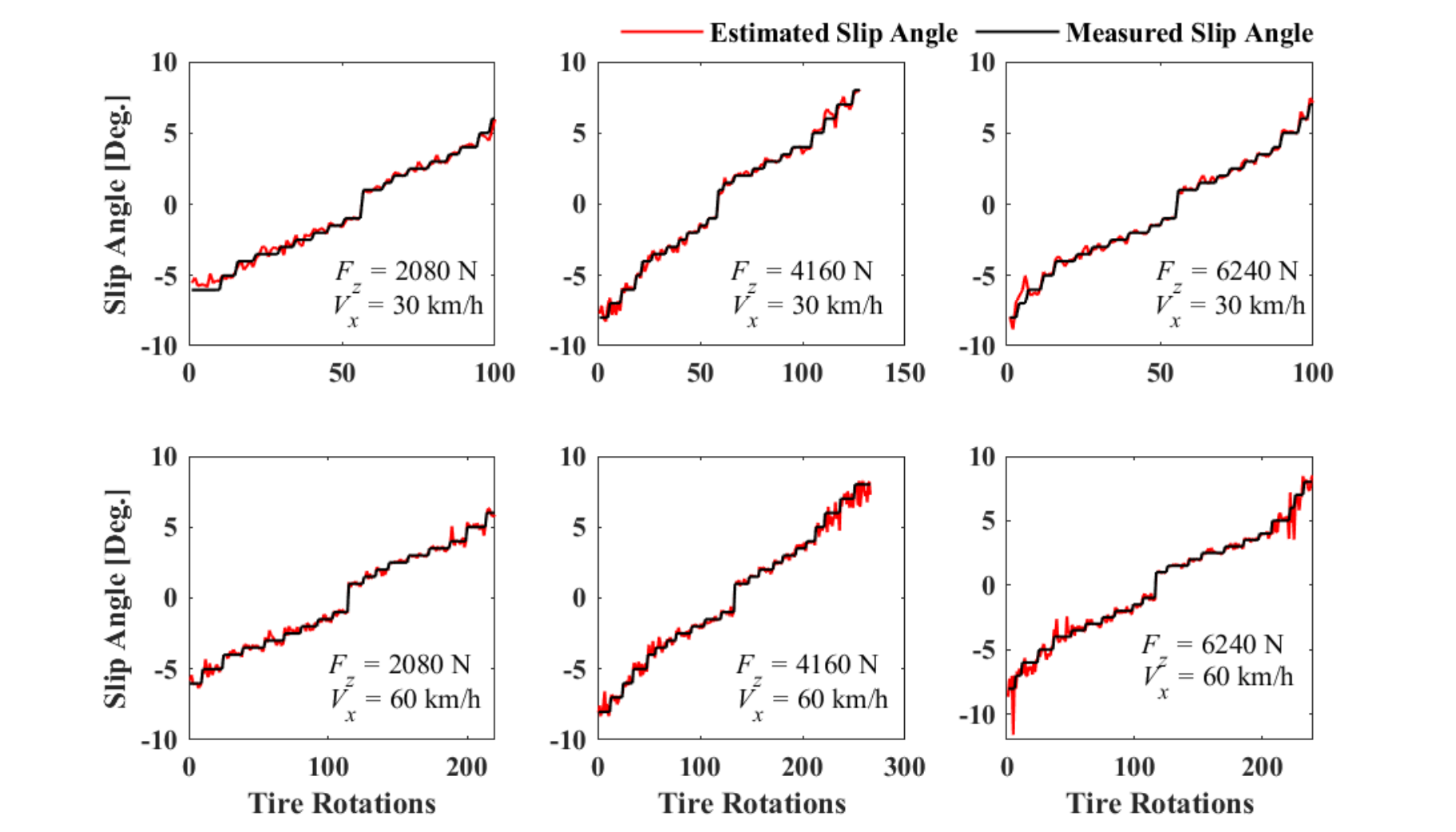}
	\caption{Estimated slip angle vs measured slip angle under step input condition}
	\label{Fig15}
\end{figure}

\begin{figure}[!htb]
	\centering
	\vspace{-1em}
	\includegraphics[width=\linewidth]{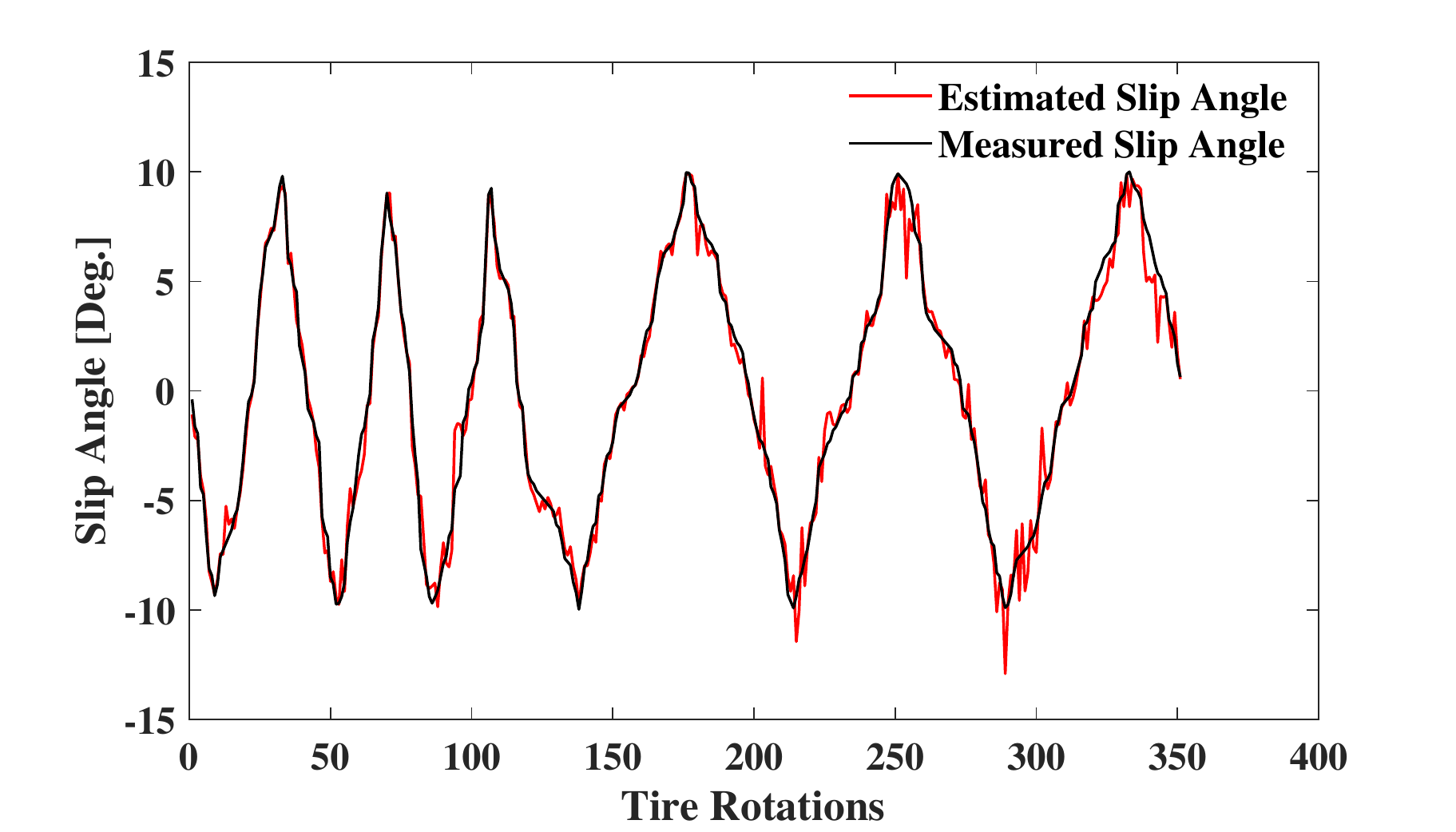}
	\caption{Estimated slip angle vs measured slip angle under triangle input condition}
	\label{Fig16}
\end{figure}

\begin{figure}[!htb]
	\centering
	\vspace{-1em}
	\includegraphics[width=\linewidth]{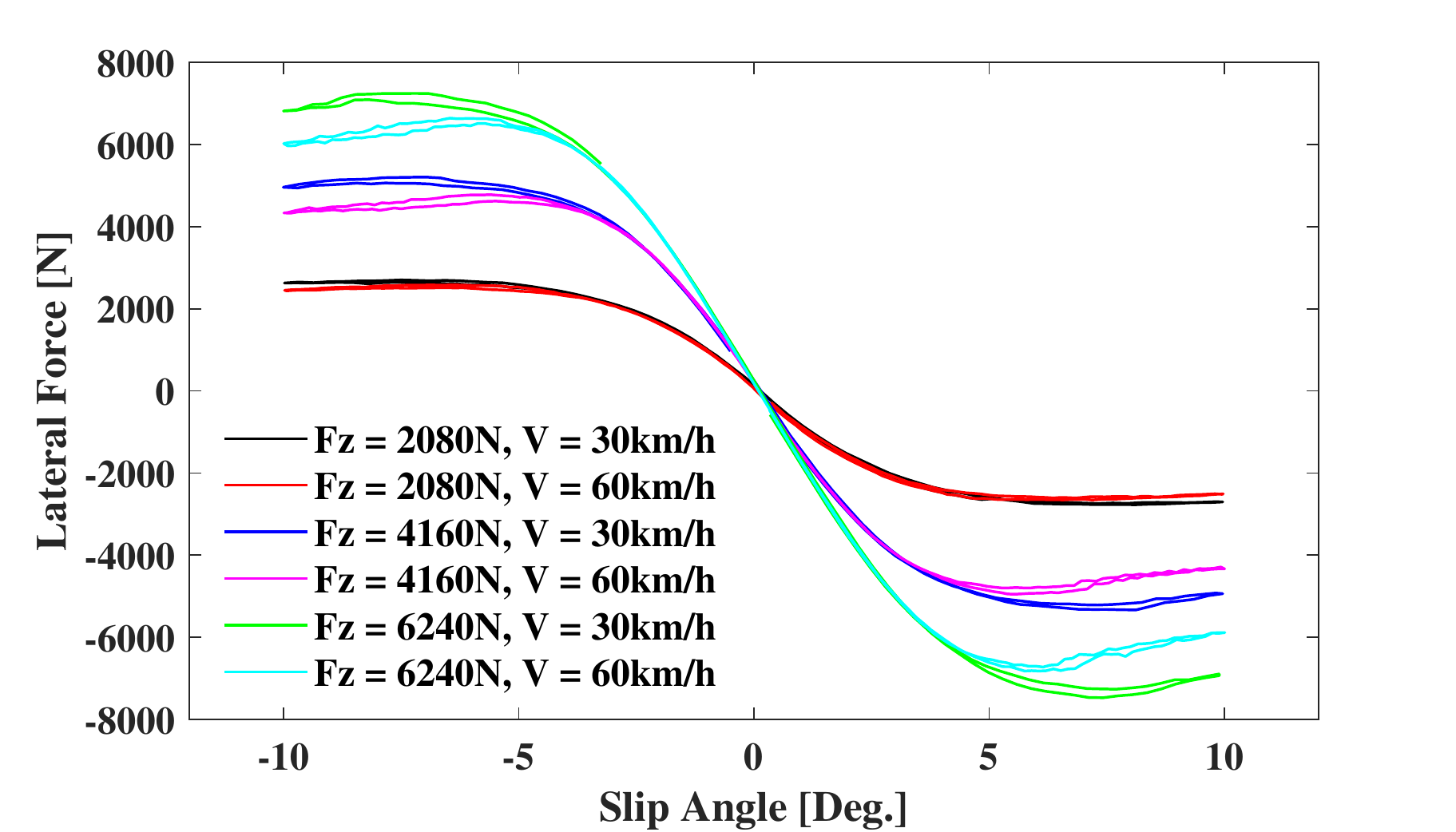}
	\caption{The tire lateral force under different experimental scenarios}
	\label{Fig17}
\end{figure}

As stated in the previous steps, the inputs are the amplitudes of 40 points that equally distribute in 0-2000 Hz range and the output is the slip angle.~The performance of slip angle estimation under the step and triangular wave input is shown in Fig. \ref{Fig15} and Fig. \ref{Fig16}. As revealed in Table \ref{Table1}, three different normal loads are applied to the tire from about 2kN to 6kN, which indicate 40 to 120 percent of the tire nominal load. Two rolling speeds (30 and 60 km/h) are provided to show the influence of velocity on smart tire.~Fig. \ref{Fig15} demonstrates the comparison of measured and estimated or predict slip angles by the proposed algorithm under the step input cases. Whereas, Fig. \ref{Fig16} shows the triangle wave input of slip angle. From these two figures, it can be observed that the maximum slip angles are up to 10 degrees, which means the tire experiences a severe sliding in the contact patch such that the lateral forces have already entered into the saturation region as shown in Fig. \ref{Fig17}. Most importantly, the results show  that the estimation algorithm can accurately estimate the slip angles. Nevertheless, some obvious errors still exist in the regions of large slip angles (i.e. 8-10 degrees) but the overall performance of the method is satisfactory.~The NRMS error is lower than 5 and 9 percents under step input and triangle wave input conditions, respectively. Furthermore, the proposed training methods or network is independent of the velocity such that it has the bright potential in real-world applications.~Compared to the existing methods, the proposed algorithm can accurately estimate the slip angle in a larger range. In addition, its estimation is based the filtered raw data instead of its integration (lateral deformation) to get rid of the stochastic errors or bias of the sensor. Furthermore, its effectiveness is verified under a more extensive scenarios compared to existing methods. 

\section{Comparison of the results between different methods}

In this section, we compare the NN based on Rprop algorithm with physical tire model and other machine learning algorithms to investigate the pros and cons of each of these algorithms for slip angle estimations.~We first focus on the physical model and show its weakness for slip angle estimation specially in harsh maneuver.~Then, a comprehensive study is presented to scrutinize the potency of Random Forest(RF) and recurrent neural network (RNN) for tire slip angle estimation in comparison with the NN based on Rprop algorithm.

\subsection{Traditional physical model}

\begin{figure}[!htb]
	\centering
	\vspace{-1em}
	\includegraphics[width=\linewidth]{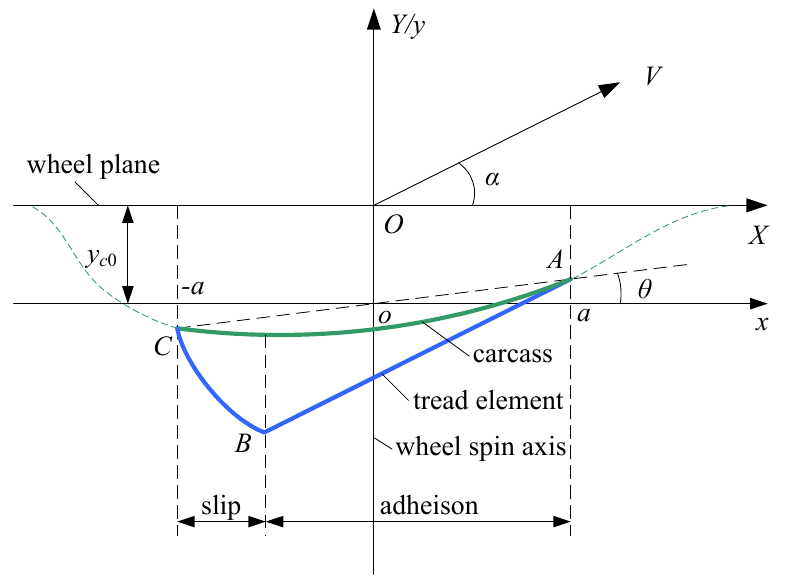}
	\caption{Deformation of tire carcass and tread under cornering conditions}
	\label{Fig18}
\end{figure}

Based on the tire modelling theory, the slip angle can be estimated by the shape of the lateral deformation curves. As shown in Fig.\ref{Fig18}, points A and C respectively represent the leading and trailing edges of the contact patch with a length of 2$a$. The blue line stands for tire tread elements and the green line represents the carcass, where the accelerometer is attached. The overall carcass deformation can be divided into translation, bending and twisting deformation. The translation deformation of the carcass is $y_{c0}$ and the twisting angle is $\theta$.~With double integrating of the acceleration signal, the lateral deformation of the carcass can be obtained. It should be noted that there is a nearly linear relationship between the slip angle and the slope at the leading edge of the carcass deformation curve.~This linear relationship can be exploited to estimate the slip angle.

\begin{figure}[!htb]
	\centering
	\vspace{-1em}
	\includegraphics[width=\linewidth]{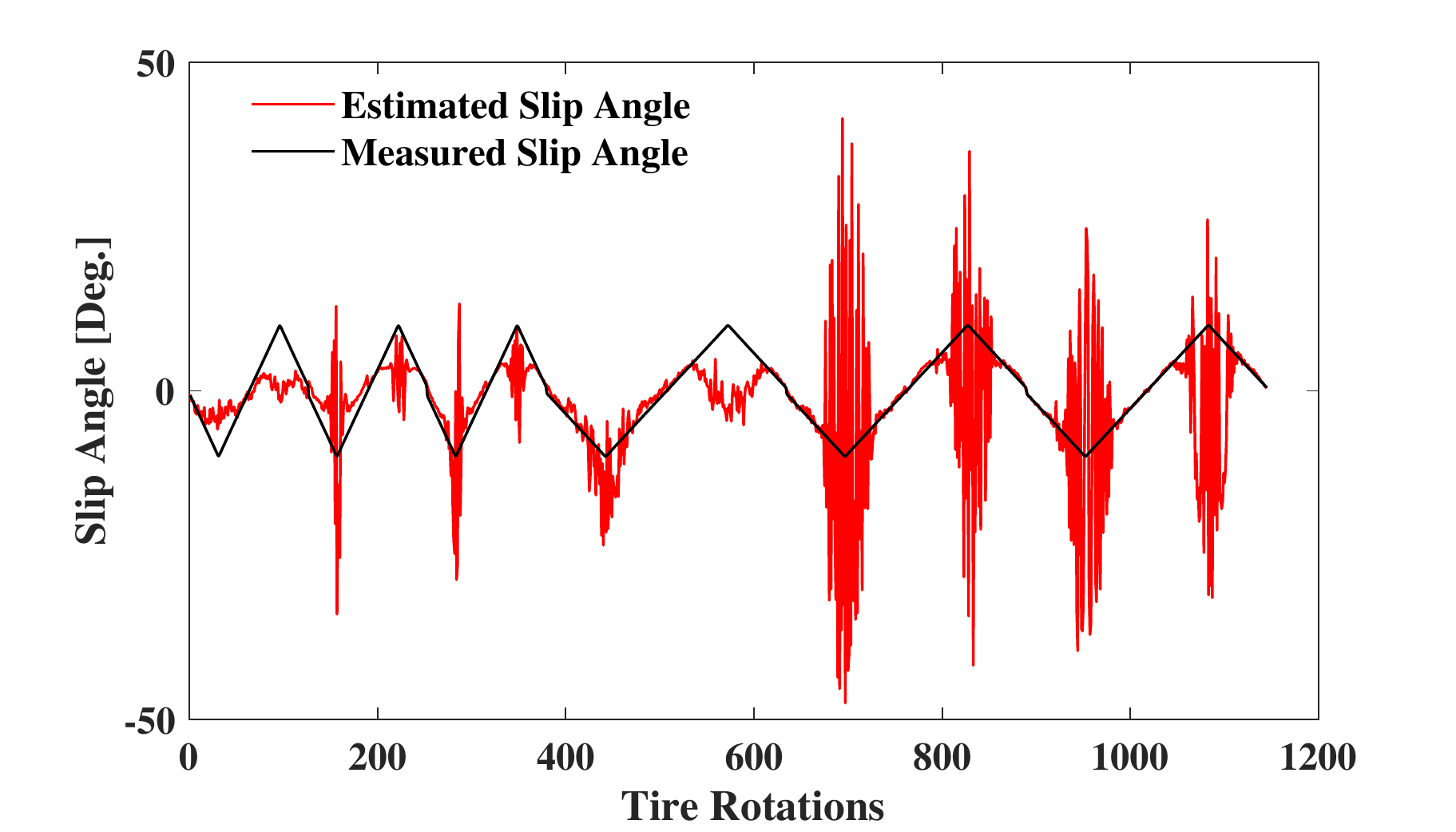}
	\caption{Estimated slip angle with analytical method}
	\label{Fig19}
\end{figure}

Based on the physical model method, the slip angle estimation results are shown in Fig.\ref{Fig19}, where the estimated slip angles within about $4^\circ$ are basically accurate, however large errors appear when the slip angle exceeds $4^\circ$.~The possible reasons are, on the one hand, the slope of the curve no longer increases after the slip angle exceeds a certain value (determined by specific tire and road status) due to the saturation of lateral force; on the other hand, the huge noises in large slip angles result in the unstable shape of lateral deformation curve and large error in the calculated slope.

\subsection{Different machine learning methods in frequency domain}

Different machine learning methods have been used in the tire modeling or parameter estimation \cite{singh2018literature}, especially the tire industry \cite{singhapplication}. In this study, two commonly used algorithms are implemented for the sake of comparison. Random Forest (RF) technique is categorized as a supervised learning method, which basically learns from labeled data and predicts in accordance with the learned patterns.~Indeed, it is an ensemble learning technique, which is mostly used for classification and regression purposes.~Generally, the RF algorithm encompasses different decision trees.~The algorithm uses and merges the decisions of several decision trees to reach an answer, which basically describes the average of all existing decisions trees\cite{ho1998random}.

The second technique used in this section is the recurrent neural network (RNN). RNN is considered as a class of neural networks, which allows the utilization of previous output as input while having hidden layers. Accordingly, RNN can implement its inside states in order to process variable length sequences of inputs. Generally, this technique  is a good fit for natural language processing and speech recognition\cite{dupond2019thorough}.

\begin{figure}[!htb]
	\centering
	\vspace{-1em}
	\includegraphics[width=\linewidth]{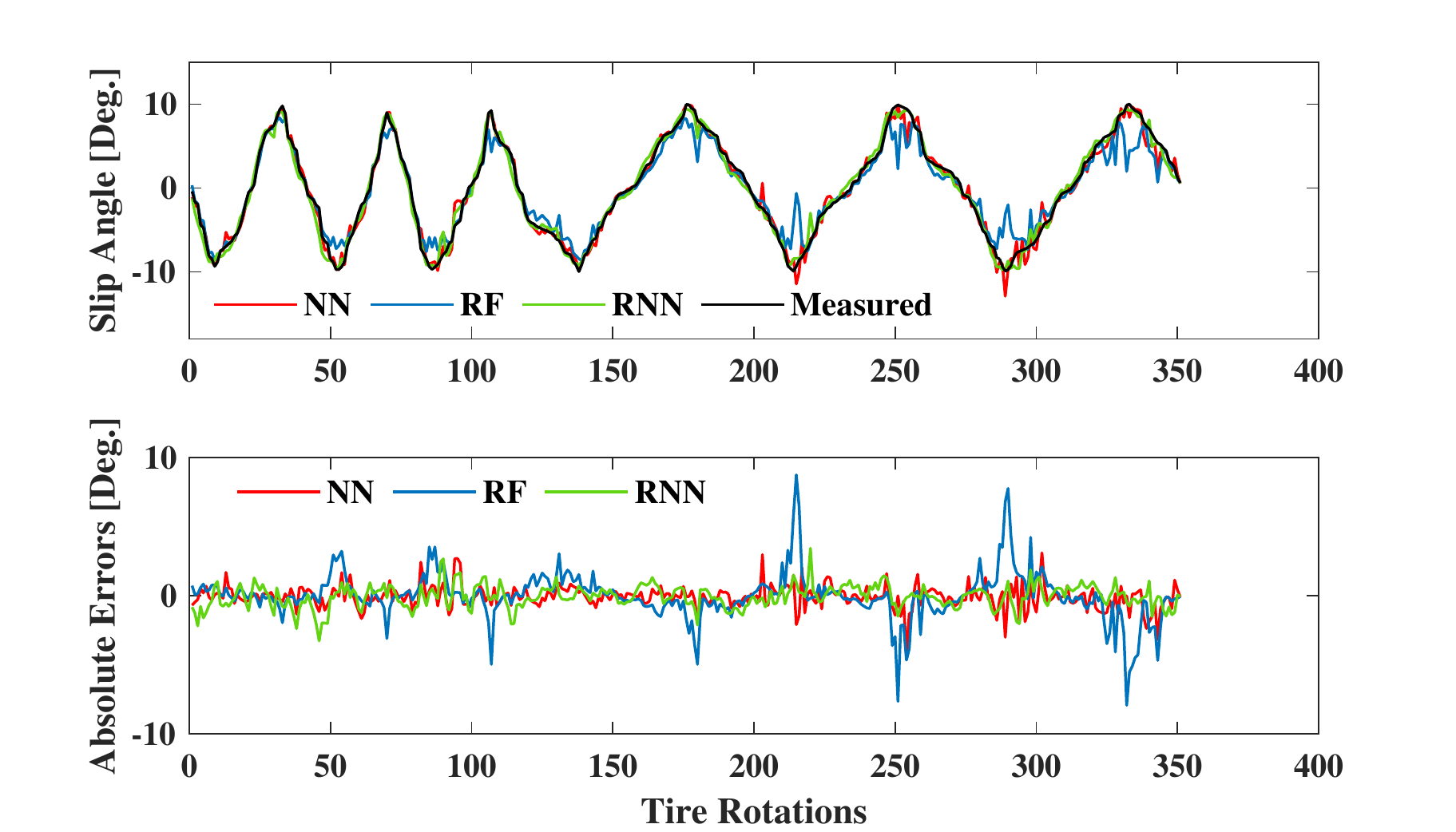}
	\caption{Comparison of the estimated results with different ML methods in frequency domain}
	\label{Fig20}
\end{figure}

We have used the same training and testing data in the frequency domain for RF and RNN algorithms, and then compared with NN based on Rprop algorithm as shown in Fig. \ref{Fig20}. As the results show, an obvious error exists at large slip angles for Random Forest, and thus, it can be concluded that this is not an effective approach for the estimation of large slip angles. However, RNN shows a good capability in tire slip angle estimations similar to NN based on Rprop algorithm. It should be noted that the RNN algorithm is more complex and need more history information as input variables (10 times data used here). For this study, it is necessary to keep the network as simple as possible in order to work in real-time applications such as vehicle safety control.~Therefore, the neural network method based on Rprop algorithm is considered as the most effective approach for tire slip angle estimation in the frequency domain.

\subsection{Different machine learning methods in time domain(400Hz, 2000Hz)}

To further delineate why the frequency data is used instead of the time-domain data, the same networks, including NN, RF and RNN are implemented to train the time-domain data. The only difference is that the inputs of the network are accelerations of 140 points along the inner liner of the tire, covering the contact patch.~The raw data is filtered by two cut-off frequencies 400Hz and 2000Hz, which correspond to low-frequency tire rolling deformation and high-frequency micro-vibration information.

\begin{figure}[!htb]
	\centering
	\vspace{-1em}
	\includegraphics[width=\linewidth]{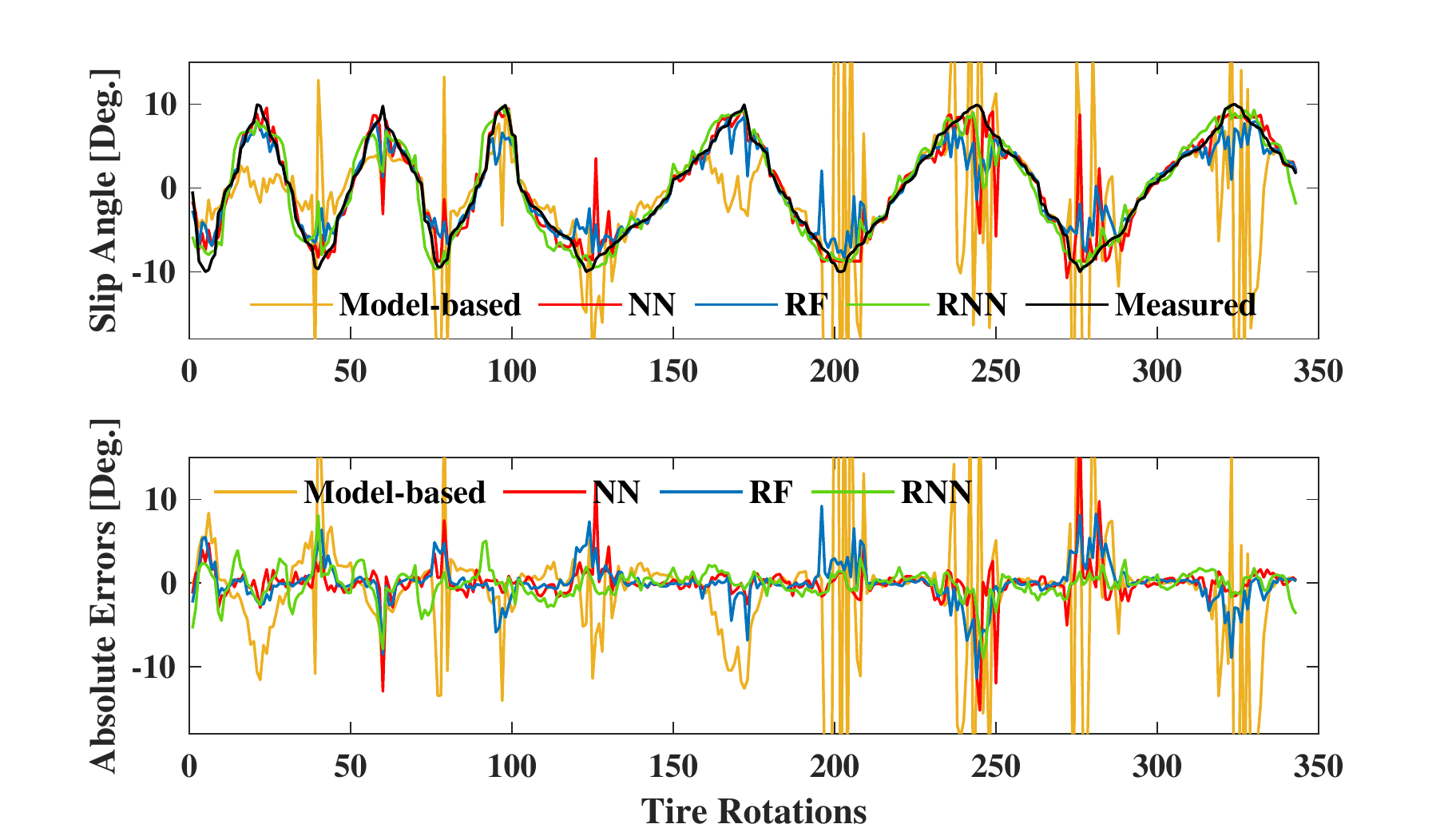}
	\caption{Comparison of estimated results with different methods in time domain (400Hz)}
	\label{Fig21}
\end{figure}

\begin{figure}[!htb]
	\centering
	\vspace{-1em}
	\includegraphics[width=\linewidth]{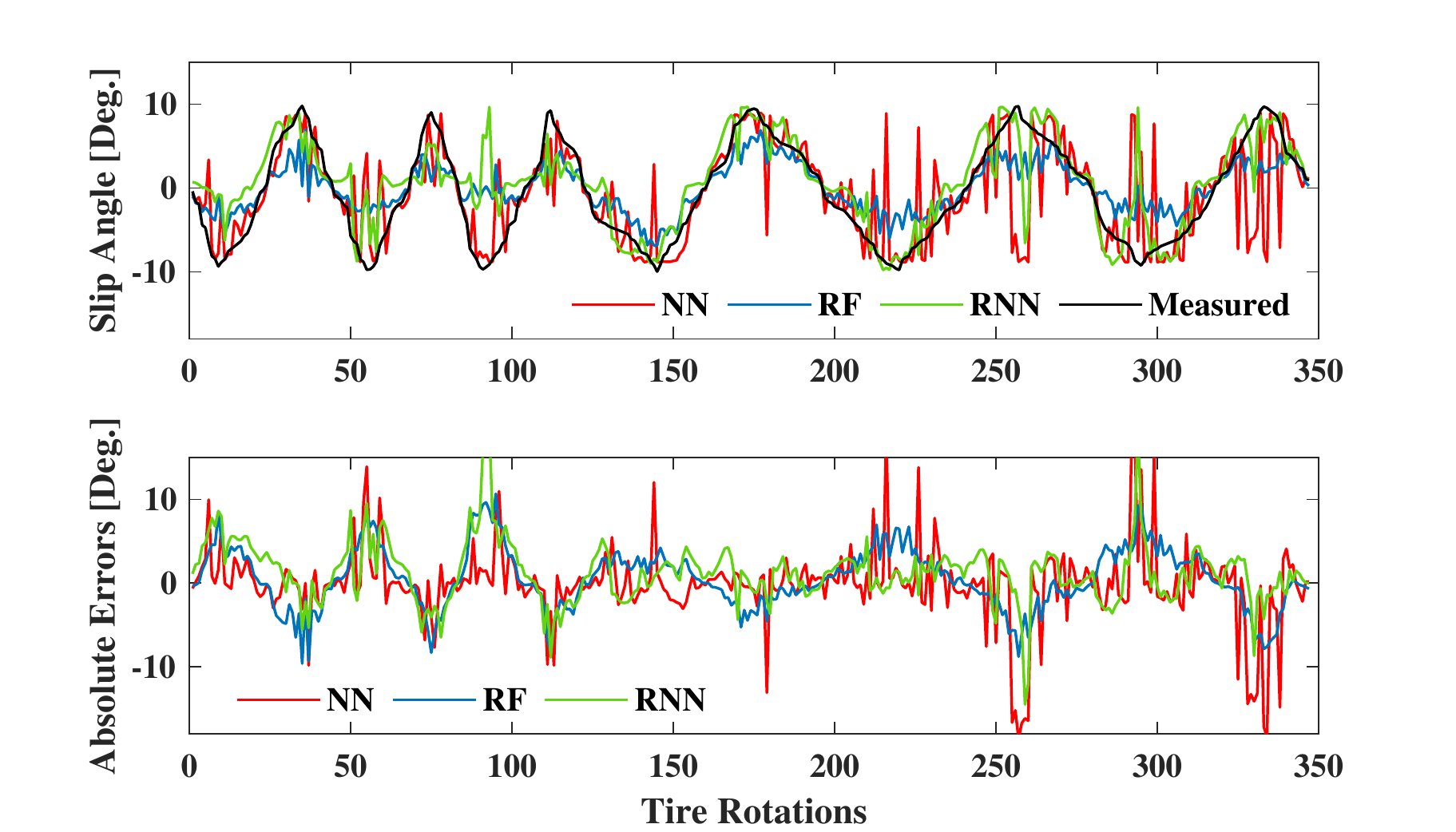}
	\caption{Comparison of estimated results with different methods in time domain (2000Hz)}
	\label{Fig22}
\end{figure}

Fig. \ref{Fig21} shows the results of different methods used in the time domain.~In the neural network based on Rprop algorithm, by using the frequency domain data, the NRMS error of the estimated slip angle is 8.33\% with only 2133 steps, however, 75073 steps are needed for training of time domain data with the NRMS error of 22.72\%. It means that the frequency domain analysis facilitates capturing the main features of the slip angle data generated by the accelerometer, which leads to higher accuracy with a faster convergence rate.~As Figs. \ref{Fig21}  and \ref{Fig22} show, the RNN algorithm with a complex network structure can provide a better performance in terms of slip angle estimation, but the improvements are still limited in comparison with the results obtained by adopting the frequency domain information.~Probably, a much better results can be obtained in time domain, if more advanced and complicated network architectures, like deep learning, are adopted to mine more valuable features.~However, in the current research, our aim is to keep the network as simple as possible with a satisfactory performance in order to make it real-time applicable if it is used in a vehicle on-board hardware. The frequency data used in this paper provides more clear features related to slip angle, which can lead to satisfactory results while retaining a simplified network.~Table \ref{Table3} provides an error analysis of different techniques utilized in the present work for tire slip angle estimation.~As the frequency domain part of the table shows, both NN based on Rprop and RNN can reach palatable amount of accuracy in terms of slip angle estimation, however, owing to simplicity of NN based on Rprop, we consider it as a better choice for real- world vehicle system applications.

\newcommand{\tabincell}[2]{\begin{tabular}{@{}#1@{}}#2\end{tabular}}
\begin{table}[htbp]
	\centering
	\caption{Error analysis of different estimation methods}\label{Table3}
	\begin{tabular}{cccc}
		\toprule
		Data used & Estimation methods & \tabincell{c}{Correlation\\ coefficient} & \tabincell{c}{NRMS\\ errors(\%)}        \\
		\midrule
		\multirow{3}{*}{\tabincell{c}{Frequency domain}} & Neural network & 0.9896 & 8.33 \\
		~ & Random forest & 0.9662 & 17.39 \\
		~ & Recurrent neural network & 0.9897 & 8.35  \\
		\midrule
		\multirow{4}{*}{\tabincell{c}{Time domain\\ (400Hz)}} & Physical model & 0.3680 & 90.88  \\
        ~ & Neural network & 0.9225 & 22.72  \\
        ~ & Random forest & 0.9306 & 23.36   \\
        ~ & Recurrent neural network & 0.9564 & 17.57   \\
        \midrule
        \multirow{3}{*}{\tabincell{c}{Time domain\\ (2000Hz)}} & Neural network & 0.6614 & 47.16   \\
        ~ & Random forest & 0.8729 & 36.61   \\
        ~ & Recurrent neural network & 0.7911 & 37.88  \\
		\bottomrule
	\end{tabular}
\end{table}

\section{Conclusion}	
This study proposed a novel estimation algorithm for tire slip angle based on the intelligent tire technology and machine-learning methods. Although the accelerometer used in the intelligent tire system can collect the accelerations in three directions, this algorithm does not require only the lateral data for the estimation. That means even the vertical or circumferential one can be also adopted to draw the similar conclusions.Then,  experiments were performed under different scenarios (e.g. different vertical local, different rolling speed, large slip angles) with step and triangle wave slip angle inputs. It was shown that the NN based on Rprop algorithm gives the highest accuracy in terms of tire slip angle estimations even with a simplified network in the frequency domain. The RNN algorithm provides satisfactory results in both frequency and time domains, however, we should consider the complexity of its network in comparison with NN based on Rprop algorithm. A comparative analysis was also presented to show the superiority of NN based on Rprop with respect to physical model estimation. The focus of future research can be on the utilization of the high-frequency data to identify the road conditions or tire noise for vehicle control purpose.

\bibliographystyle{IEEEtran}
\bibliography{mybib}	

\begin{IEEEbiography}
[{\includegraphics[width=1in,height=1.2in,clip,keepaspectratio]{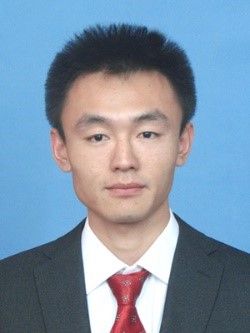}}]{Nan Xu} received the Ph.D. degree in vehicle engineering from Jilin University, Changchun, China, in 2012. He is currently an associate professor at State Key Laboratory of Automotive Simulation and Control, Jilin University. His current research focuses on tire dynamics, intelligent tire, vehicle dynamics, stability control of electric vehicles and autonomous vehicles.
\end{IEEEbiography}
\vspace{3 cm}

\begin{IEEEbiography}[{\includegraphics[width=1in,height=1.25in,clip,keepaspectratio]{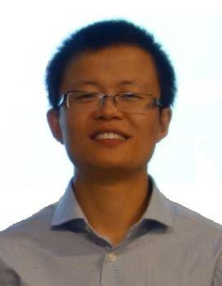}}]{Yanjun Huang} is a Research Associate at the Department of Mechanical and Mechatronics Engineering at University of Waterloo, where he received his PhD degree in 2016. His research interest is mainly on the vehicle holistic control in terms of safety, energy-saving, and intelligence, including vehicle dynamics and control, HEV/EV optimization and control, motion planning and control of connected and autonomous vehicles, human-machine cooperative driving. 
He has published several books, over 60 papers in journals and conference; He is the recipient of IEEE Vehicular Technology Society 2019 Best Land Transportation Paper Award, the 2018 Best paper of Automotive Innovation, and top 10 most popular paper in Journal Automobile Technology. He is serving as associate editors and editorial board member of IET Intelligent Transport System, SAE Int. J. of Commercial vehicles, Int. J. of Autonomous Vehicle system, etc.
\end{IEEEbiography}
\vspace{-4 cm}
\begin{IEEEbiography}[{\includegraphics[width=1in,height=1.25in,clip,keepaspectratio]{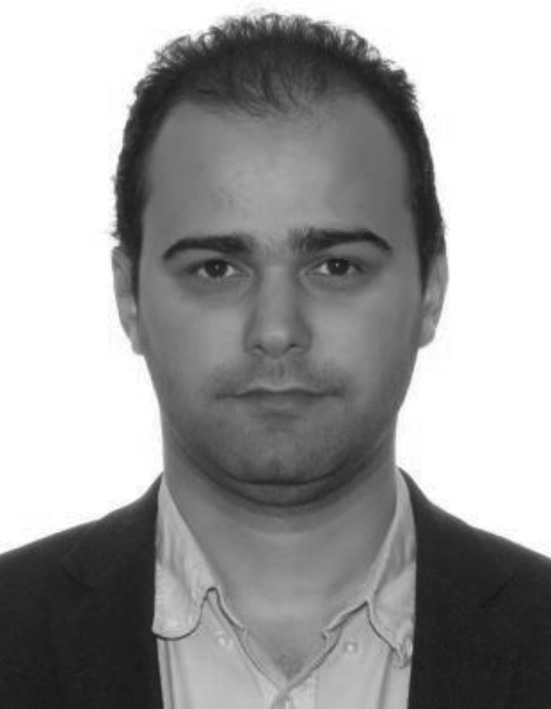}}]{Hassan Askari} was born in Rasht, Iran and received his B. Sc. ,M.Sc. and PhD degrees from Iran University of Science and Technology, Tehran, Iran,   University of Ontario Institute of Technology, Oshawa, Canada, and  University of Waterloo, Waterloo, Canada in 2011,  2014, and 2019  respectively. He published more than 70 journal and conference papers in the areas of nonlinear vibrations, applied mathematics,  nanogenerators and self-powered sensors. He co-authored one book and one book chapter both published by Springer. He is an active reviewer for more than 40 journals and editorial board member of several scientific and international journals. He has received several prestigious awards including, Outstanding Researcher at the Iran University of Science and Technology, Fellowship of the Waterloo Institute of Nanotechnology, NSERC Graduate Scholarship, Ontario Graduate Scholarship, and the University of Waterloo President Award. He was nominated for the Governor General's Academic Gold Medal at the University of Ontario Institute of Technology and University of Waterloo in 2014 and 2019, respectively. He is currently a Postdoctoral Fellow at the Department of Mechanical and Mechatronics Engineering at the University of Waterloo.
\end{IEEEbiography}
\vspace{-3.5 cm}
\begin{IEEEbiography}[{\includegraphics[width=1in,height=1.25in,clip,keepaspectratio]{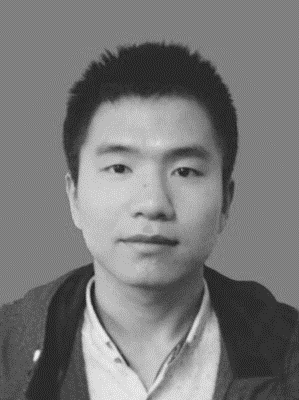}}]{Zepeng Tang} is currently a M.S. candidate in the College of Automotive Engineering, Jilin University, Changchun, China. His research interest is mainly on intelligent tire, vehicle dynamics and autonomous vehicles.
\end{IEEEbiography}	
\vspace{3.5 cm}

\end{document}